
\input phyzzx.tex


\def\np{Nucl. Phys.}
\def\pl{Phys. Lett.}
\def\prl{Phys. Rev. Lett.}

\def\cmp{Comm. Math. Phys.}
\def\ijmp{Int. J. Mod. Phys.}
\def\mpl{Mod. Phys. Lett.}

\def\phyrep{Phys. Rep.}

\tolerance=500000
\overfullrule=0pt
\Pubnum={US-FT-5/95}
\pubnum={US-FT-5/95}
\date={May, 1995}
\pubtype={}
\titlepage

\title{TOPOLOGICAL MATTER, MIRROR SYMMETRY AND NON-CRITICAL
(SUPER)STRINGS }

\author{A.V. Ramallo
\foot{E-mail: ALFONSO@GAES.USC.ES}
 and J.M. Sanchez de Santos\foot{E-mail: SANTOS@GAES.USC.ES}}
\address{Departamento de F\'\i sica de
Part\'\i culas, \break Universidad de Santiago, \break
E-15706 Santiago de Compostela, Spain.}

\abstract{We study the realization of the (super)
conformal topological symmetry in  two-dimensional field theories.
The mirror automorphism of the topological algebra is represented as
a reflection in the space of fields. As a consequence, a double BRST
structure for topological matter theories is found. It is shown that
the implementation of the topological symmetry in non-critical
(super)string theories depends on the matter content of the two
realizations  connected by the mirror transformation.}

\endpage
\pagenumber=1
\hyphenation {o-pe-ra-tor}

\chapter{Introduction}

Two-dimensional topological field theories have been a subject
extensively studied in the last few years
\REF\wittop{E. Witten \journal\cmp&117(88)353.}
\REF\wit{E. Witten \journal\cmp&118(88)411.}
\REF\bbrt{For a review see D. Birmingham, M. Blau, M.Rakowski
and G. Thompson \journal\phyrep&209(91)129.}[\wittop, \wit,
\bbrt]. These models are endowed with a topological BRST symmetry that
allows to eliminate all their local excitations. If, in addition to
the topological symmetry, the system is conformally invariant, we say
that the model is a topological conformal field theory (TCFT)
\REF\dij{R. Dijkgraaf, E. Verlinde and H. Verlinde
\journal\np&B352(91)59.;``Notes on topological string theory
and 2d quantum gravity", Proceedings of the Trieste spring
school 1990, edited by M.Green et al. (World Scientific,
Singapore,1991).}[\dij].

As any other
conformal field theory, the TCFT's are characterized by their
chiral algebra, which must include the Virasoro algebra as a
subalgebra.  The topological nature of TCFT's is ensured if all the
generators of the chiral algebra (including the energy-momentum
tensor) are exact in the cohomology defined by the   topological
BRST charge.

The compatibility between the topological and the (extended)
conformal symmetries is encoded in the so-called topological algebra
[\dij].
 This is the operator algebra closed by the BRST current and the
generators of the chiral algebra. The realization of the topological
algebra in different topological matter systems is the main subject
studied in this paper.

The standard procedure to obtain the local form of the topological
symmetries corresponding to some chiral algebras consists in performing
a redefinition (a twist) of different superconformal theories. For
example, the local BRST algebra for the (unextended) Virasoro symmetry
can be obtained from the $N=2$ superconformal models
\REF\LVW{W. Lerche, C. Vafa and N.P. Warner
\journal\np&B324(89)427.}
\REF\EY{T. Eguchi and S.-K. Yang \journal\mpl&A4(90)1653;
T. Eguchi, S. Hosono and S.-K. Yang \journal\cmp&140(91)159.}[\LVW,
\EY],
 whereas if the
topological matter system is ($N=1$) supersymmetric, its local BRST
symmetry is governed by a twisted $N=3$ superconformal algebra
\REF\yos{H. Yoshii \journal \pl&B259(91)279.}[\yos]. In
general the twisting procedure gives rise to a Virasoro algebra with a
vanishing central charge. Nevertheless, the twisted algebra contains
c-number anomalies, parametrized by a number (the so-called
dimension) which characterizes the TCFT.

One of the features of the topological algebra that we shall study is the
fact that it possesses an automorphism, the mirror transformation
\REF\Mirror{For a review of the applications of mirror symmetry to
the study of complex manifolds see ``Essays on Mirror
Manifolds", edited by S. T. Yau (International Press, Hong Kong,
1992).}[\Mirror],
that relates two different realizations. Under this mirror
transformation the abelian currents needed to implement locally the
topological symmetry  change their signs and the BRST
current is exchanged with the BRST ancestor of the energy-momentum
tensor. The origin of the mirror symmetry can be traced back to the
twisting procedure since the two representations of the topological
algebra connected by a mirror transformation are obtained when two
different twisting prescriptions are applied to the same
superconformal field theory.
Within our general approach, the mirror transformation will
be represented as a transformation of fields. In fact, when we are in
a convenient basis of fields, we shall see that the mirror symmetry
can be understood as a reflection in field space.

The topological conformal symmetry is realized in systems where
conformal matter is coupled to $2d$ (super)gravity
\REF\kpz{A.M. Polyakov \journal\mpl&A2(87)893; V. Knizhnik,
A.M. Polyakov and A.B. Zamolodchikov \journal\mpl&A3(88)819.}
\REF\DDK{F. David \journal\mpl&A3(88)1651; J. Distler and H. Kawai
\journal\np&B321(89)509.}
\REF\gins{For a review see P. Ginsparg and G. Moore,
``Lectures on 2D Gravity and  2D String
Theory"(hep-th/9304011), in `` Recent directions in particle
Theory", Proceedings of the 1992 TASI summer school, edited by J.
Harvey and J. Polchinski (World Scientific, Singapore, 1993).}[\kpz,
\DDK, \gins],  as  happens in
non-critical (super)string theories.
Indeed, as  was first
shown in
\REF\gato{B. Gato-Rivera and A.M.
Semikhatov\journal\pl&B293(92)72 \journal\np&B408(93)133.}
ref. [\gato], one can improve the usual BRST current that
fixes the reparametrization invariance in such a way that it
realizes the topological algebra. Moreover, this result has been
extended in
\REF\BLNW{M. Bershadsky, W. Lerche, D. Nemechansky and N.
Warner\journal\np&B401(93)304.}[\BLNW] to systems where matter is
coupled to supergravity and $W$-gravity (see also ref.
\REF\Mukhi{S. Mukhi and C. Vafa\journal\np&B407(93)667.}[\Mukhi]).

In these realizations of the
topological symmetry, the field content is given by matter (and
Liuoville) fields together with (super)diffeomorphism ghosts. The
ghost sector in these matter+gravity models always contains a pair
 $(b,c)$ of anticommuting ghosts with spins $(2,-1)$. By means of
the mirror transformation, the ghosts $b$ and $c$ acquire spins $1$
and $0$ respectively, whereas the matter central charge is changed.
For these systems we shall find that in order to have a
representation of the topological algebra with a real c-number
extension (i.e. with real dimension) the central charges of the
matter sectors of the two realizations of the topological symmetry
connected by the mirror transformation must be restricted to a
particular range of values. Actually we shall obtain a ``barrier"
formula which, surprisingly, involves the two
realizations of the topological symmetry related by the mirror
reflection.

This paper is organized as follows. In section 2 we study the
topological symmetry in a system of scalar fields with vanishing
Virasoro central charge. After introducing the topological algebra,
we study its possible vertex operator representations. By means of a
convenient bosonization, the results obtained are put in terms of a
pair of fermionic fields which, after a mirror transformation, are
identified with the diffeomorphism ghosts. This section concludes
with a study of the reflections in field space that are equivalent to
the mirror automorphism. The realization of the topological symmetry
in the bosonic string theory is analysed in section 3. In this
section we explore the role played by the mirror invariance in the
implementation of the topological symmetry in the models of matter
coupled to two-dimensional gravity.

In section 4 we analyze a supersymmetric model of scalar bosons and
Majorana fermions. The supersymmetric extension of the topological
algebra is reviewed at the beginning of this section. In order to get
a representation of this algebra in our model, a supersymmetric
ghost system is introduced. In terms of these ghosts fields the
topological supersymmetric algebra is easily realized. In one of
these realizations one can identify the  ghost fields with
the standard superdiffeomorphism ghosts  of two-dimensional
supergravity. Many of the results obtained in sections 2
and 3 are generalized to the supersymmetric case. In particular, the
mirror transformation is represented as a simultaneous reflection
of the scalar and Majorana fields. Moreover, the range of values of
the central charge of the matter sector required to implement the
topological symmetry with a real dimension can be determined as in
the non-supersymmetric case.

We summarize our results and discuss the possible extensions of our
work in section 5. Finally in the Appendix we give some details of our
calculations.

\chapter{The BRST symmetry of topological matter}

Let us consider a system of conformal topological matter in two spacetime
dimensions. Such a system is characterized by an energy-momentum tensor $T$
that closes the Virasoro algebra with vanishing central charge. Therefore
the holomorphic component of $T$ satisfies the operator product expansion
(OPE) :
$$
T(z)T(w)={2T(w)\over (z-w)^2} +{\partial T(w)\over z-w}.
\eqn\uno
$$
We shall consider a realization of the algebra \uno\ in terms of a
multicomponent scalar field $\vec \phi(z)=(\phi_1(z),\ldots,\phi_N(z))$, $N$
being the number of scalar fields. The $\phi_i$'s are
free fields and therefore they  obey the basic OPE's
$$
\phi_i(z)\,\,\phi_j(w)=-\delta_{ij}\,\,log(z-w).
\eqn\dos
$$
In terms of $\vec \phi$, the energy-momentum tensor $T$ can be written as :
$$
T=-{1\over 2}\partial\vec\phi\cdot\partial\vec\phi+
\vec A\cdot\partial^2\vec\phi,
\eqn\tres
$$
where we have included some background charges parametrized by the constant
vector $\vec A$ ($\vec A \in {\bf C}^N$). For a general  $\vec A$, $T(z)$
closes the Virasoro algebra with central charge
$$
c=N+12\vec A\,^{2}.
\eqn\cuatro
$$
Therefore, in order to satisfy eq. \uno, we must require  $c$ to vanish
and, as a consequence, this fixes the value of $\vec A\,^{2}$ to be :
$$
\vec A\,^{2}=-{N\over 12}.
\eqn\cinco
$$

We would like this model to be a topological theory. This would be the case
if we were able to find a nilpotent BRST current $Q$ such that it could
be considered as the generator of a topological symmetry. The nilpotency
of $Q$ is guaranteed if the OPE of $Q$ with itself is regular :
$$
Q(z)\,Q(w)=0,
\eqn\seis
$$
whereas the requirement of being the generator of a symmetry is
fulfilled if we impose that $Q$ be a primary dimension-one operator with
respect to the energy-momentum tensor $T$ :
$$
T(z)\,Q(w)={Q(w)\over (z-w)^2} +{\partial Q(w)\over z-w}.
\eqn\siete
$$
What makes $Q(z)$ the generator of a topological symmetry is the fact that
the energy-momentum tensor $T$ is $Q$-exact, \ie\ that there exists an
operator $G$ such that
$$
T(z)=\,\,\{\oint Q,G(z)\}.
\eqn\ocho
$$
$G$ is thus a dimension-two operator which is the BRST partner of $T$. On
dimensional grounds, the local version of eq. \ocho\ has the form :
$$
Q(z)\,G(w)={d\over (z-w)^3}+ {R(w)\over (z-w)^2} +{T(w)\over z-w},
\eqn\nueve
$$
where $d$ is a c-number constant and $R$ is an abelian dimension-one current.
We shall call the algebra closed by $Q$, $T$, $G$ and $R$, the topological
algebra (TA). This algebra characterizes the topological symmetry of the
model. In its minimal version the algebra closes without introducing extra
generators. Consistency with eqs. \seis, \siete\ and \nueve\ is achieved if
one requires the other OPE's of the algebra to be:
$$
\eqalign{
R(z)\,R(w)=&{d\over (z-w)^2}\cr
R(z)\,Q(w)=&{Q(w)\over z-w}\cr
T(z)\,R(w)=&-{d\over (z-w)^3}+{R(w)\over (z-w)^2}+
{\partial R(w)\over z-w}\cr
R(z)\,G(w)=&-{G(w)\over z-w}\cr
T(z)\,G(w)=&{2G(w)\over (z-w)^2}+{\partial G(w)\over z-w}\cr
G(z)\,G(w)=&0.\cr}
\eqn\diez
$$
The central extension $d$ of the algebra will be called the dimension of the
TA [\dij]. Notice that $R$  is an anomalous abelian current whose
anomaly is precisely given by $d$. In fact the TA displayed in eqs.
\uno, \seis,  \nueve\ and \diez\ can be obtained by twisting the $N=2$
superconformal algebra. The relation between the parameter $d$ and
the Virasoro central charge of the $N=2$ theory is simply
$c^{N=2}=3d$. An interesting aspect of the TA is that for any
realization of the algebra one can generate another with the same
value of $d$ by means of a redefinition that exchanges the role of
$Q$ and $G$ and changes the sign of $R$ : $$
\eqalign{
T\rightarrow T^{*}=&\,\,T-\partial R\cr
Q\rightarrow Q^{*}=&\,\,G\cr
G\rightarrow G^{*}=&\,\,Q\cr
R\rightarrow R^{*}=&-R.\cr}
\eqn\once
$$
The transformation in \once\ will be called the mirror transformation. It
has its origin in the two possible twistings $T^{N=2}\pm {1\over 2}
\partial R$ that one can perform to generate a topological theory from a
$N=2$ superconformal model.

Let us now investigate the possible realizations of the BRST current in our
scalar field theory. We want $Q$ to be a primary dimension-one operator
local in the field $\vec \phi$. Accordingly we shall adopt the following
ansatz for $Q$ :
$$
Q=Q_n(\partial^{i}\phi)\,\,e^{\vec \alpha\cdot\vec \phi},
\eqn\doce
$$
where $Q_n$ is a polynomial in the derivatives of $\vec\phi$ with dimension
$n$. This non-negative integer $n$ will be called the depth of the operator
$Q$.
In eq. \doce\ $\vec \alpha$ is a constant vector. The nilpotency condition
of $Q$ (eq.\seis) implies that:
$$
\vec \alpha\,^{2} \leq 0.
\eqn\trece
$$
On the other hand, as the conformal weight with respect to the
energy-momentum tensor $T$ in eq. \tres\ of the vertex operator
$e^{\vec \alpha\cdot\vec \phi}$ is
$$
\Delta (e^{\vec \alpha\cdot\vec \phi})=
\vec A\cdot\vec \alpha-{1\over 2}\vec\alpha\,^{2},
\eqn\catorce
$$
it follows from eq. \siete\ that
$$
\Delta_{Q}=n+\vec A\cdot\vec \alpha-{1\over 2}\vec\alpha\,^{2}=1.
\eqn\quince
$$
Once $Q$ is fixed, one must determine an operator $G$ such that eq. \ocho\
holds. In view of the ansatz we have taken for $Q$, it is natural to search
for an operator $G$ of the same form, \ie\ a polynomial $G_m$ of depth $m$
times a vertex operator :
$$
G=G_m(\partial^{i}\phi)\,\,e^{-\vec \alpha\cdot\vec \phi}.
\eqn\dseis
$$

Notice that we are forced to have the exponential factor
$e^{-\vec \alpha\cdot\vec \phi}$ in $G$ if we want to fulfill eq. \ocho\
for $T$ as in eq. \tres. The dimension-two character of $G$ constraints
$m$, $\vec A$, and $\vec \alpha$ to satisfy :

$$
\Delta_{G}=m-\vec A\cdot\vec \alpha-{1\over 2}\vec\alpha\,^{2}=2.
\eqn\dsiete
$$
Adding eqs. \quince\ and \dsiete, one gets a simple equation relating the
depths of $Q$ and $G$ with the length of the vector $\vec \alpha$ :
$$
n+m=\vec \alpha\,^{2}+3.
\eqn\docho
$$
As  $\vec\alpha\,^{2}$ cannot be positive (see eq. \trece),  we
get from eq.\docho\ the following restriction on the possible values of the
depths :

$$
n+m\leq 3.
\eqn\dnueve
$$
Moreover, by substracting eqs. \quince\ and \dsiete, the scalar product
$\vec A \cdot\vec \alpha$ is determined as a function of $n$ and $m$ :
$$
\vec A\cdot\vec\alpha={m-n-1\over 2}.
\eqn\veinte
$$
The rule \veinte\ greatly restricts the number of possible choices for
$n$ and $m$. Actually the mirror symmetry of the algebra pairs the
$(n,m)$ and $(m,n)$ cases since by applying a mirror transformation to a
$(n,m)$ realization one generates a $(m,n)$ one :
$$
(n,m)\,^{*}\,\approx\,(m,n).
\eqn\vuno
$$
Up to now we have imposed  very few conditions extracted from the TA. In
order to implement the topological symmetry one must check the closure of the
full algebra. This has to be done case by case for the different values of
the depths $n$ and $m$.  It turns out that only for
$(n,m)=(0,2)$ and its mirror image $(m,n)=(2,0)$ the algebra can be closed
in general for $N\geq 2$.
  For other values of $n$ and $m$ the constraints generated by the
algebra are either inconsistent or they can only be solved for particular
values of the number $N$ of scalar fields. The analysis of the cases allowed
by the rule \dnueve\ is presented in the Appendix. In the present section we
shall limit ourselves to studying the $(0,2)$ and $(2,0)$ cases. Notice
that according to eq. \veinte\ when $n=0$ and $m=2$ one has:
$$
\vec \alpha^{2}=-1
\,\,\,\,\,\,\,\,\,\,\,\,\,\,
\vec A\cdot\vec\alpha={1\over 2}.
\eqn\vdos
$$
Let us parametrize $Q$ and $g$ in this case as follows:
$$
\eqalign{
Q=&\,e^{\vec \alpha\cdot\vec \phi}\cr
G=&(\vec g\cdot \partial ^2\vec\phi +\partial\vec\phi\cdot H\cdot
\partial\vec\phi) e^{-\vec \alpha\cdot\vec \phi},\cr}
\eqn\vtres
$$
where $\vec g$ is a constant vector and $H$ is a numerical $N\times N$
symmetric matrix. In the following we shall denote by $\vec
X\cdot H \cdot\vec Y$ to the contraction of the matrix $H$ with any
two vectors $\vec X$ and $\vec Y$. In order to identify $d$ and $R$
one has to compute the OPE of $Q$ and $G$ (see eq. \nueve). A
straightforward calculation leads to the result:
$$ \eqalign{
d=&\vec g\cdot\vec\alpha+\vec\alpha\cdot H\cdot\vec\alpha\cr
R=&(2\vec\alpha\cdot H+d\,\vec\alpha)\cdot\partial\vec\phi\equiv\vec
R\cdot\partial\vec\phi.\cr}
\eqn\vcuatro
$$
On the other hand, by comparing the background charge and kinetic terms of
$T$ with those appearing in the simple pole singularity of the product
$Q(z)G(w)$, one gets the constraints :
$$
\eqalign{
\vec A=&\vec g+{d\over 2}\vec\alpha\cr
{d \over 2}\,\vec
\alpha\otimes\vec\alpha&+(H\cdot\vec\alpha)\otimes\vec\alpha+
\vec\alpha\otimes(H\cdot\vec\alpha)+H=-{I\over 2},\cr}
\eqn\vcinco
$$
where $I$ is the $N\times N$ identity matrix. The anomalous character of the
$R$-current is reflected in the double and triple pole singularities
appearing in the OPE's $R(z)R(w)$ and $T(z)R(w)$. Parametrizing $R=\vec
R\cdot \partial\vec\phi$, with $\vec R$ as given in the second equation in
\vcuatro, one obtains two new conditions
$$
\vec A\cdot\vec R=-{d\over 2}
\,\,\,\,\,\,\,\,\,\,\,\,\,\,\,
\vec R\cdot\vec R=-d.
\eqn\vseis
$$
Other OPE's of the TA follow from the conditions we have obtained so far
and therefore one can consider eqs. \vcuatro, \vcinco\ and \vseis\ as a
complete set of consistency conditions that ensure the fulfillment of
the TA. In
fact one can use eqs. \vcuatro -\vseis\ to get new relations between $H$,
$\vec \alpha$ and $\vec g$. For instance, if we contract both sides of the
second equation in \vcinco\ with two vectors $\vec \alpha$, we get :

$$
\vec\alpha\cdot H\cdot\vec\alpha={d-1\over 2}.
\eqn\vsiete
$$
Plugging this result in eqs. \vcuatro\ and \vseis, one can readily obtain :
$$
\vec g\cdot\vec\alpha={d+1\over 2}
\,\,\,\,\,\,\,\,\,\,\,\,\,\,\,\,
\vec \alpha\cdot H\cdot\vec g =-{d(d+1)\over 2}.
\eqn\vocho
$$

Let us check that the constraints found for $H$, $\vec \alpha$ and $\vec g$
are enough to ensure that $Q$ and $G$ have good quantum numbers with
respect to the $R$  current. A direct calculation using eqs. \vtres\ and
\vcuatro\ yields the result :
$$
R(z)\,Q(w)=-{\vec R\cdot\vec \alpha\over z-w}\,
e^{\vec \alpha\cdot\vec \phi(w)}.
\eqn\vnueve
$$
Using eqs. \vdos\ and \vsiete\ one immediately finds for the scalar
product  $\vec R \cdot\vec\alpha$ the value
$$
\vec R\cdot\vec \alpha =-1,
\eqn\treinta
$$
from which the OPE $R(z)Q(w)$ of eq. \diez\ follows. Similarly, after a
simple calculation, one gets :
$$
R(z)\,G(w)=-{2\vec R\cdot\vec g\over (z-w)^3}\,
e^{-\vec \alpha\cdot\vec \phi(w)} -\,\,
{2\vec R\cdot H\cdot\partial \vec \phi(w)\over
(z-w)^2}\,e^{-\vec \alpha\cdot\vec \phi(w)}
+\vec R\cdot\vec \alpha \,\,{G(w)\over z-w}.
\eqn\tuno
$$
It is easy to prove that the residues of the triple and double poles in eq.
\tuno\ vanish. Indeed, making use of eqs. \vcinco, \vsiete\ and \vocho, one
easily arrives at :
$$
\vec R\cdot\vec g=0
\,\,\,\,\,\,\,\,\,\,\,\,\,\,\,\,
H\cdot\vec R=0,
\eqn\tdos
$$
which, together with eq. \treinta, imply that $R(z)$ and $G(w)$ have the OPE
displayed in eq. \diez. Proceeding in the same way,  all the
remaining OPE's in eq. \diez\ can be checked.

The realization of the TA we have obtained can be put in much simpler terms
if one introduces a pair $(b,c)$ of two anticommuting fields. In terms of
the bosonic field $\vec \phi$, these new fields are given by standard
bosonization formulas that involve the component of $\vec \phi$ along the
direction of $\vec \alpha$:
$$
b=e^{\vec \alpha\cdot\vec\phi}
\,\,\,\,\,\,\,\,\,\,\,\,\,
c=e^{-\vec \alpha\cdot\vec\phi}.
\eqn\ttres
$$
Notice that from eq. \vdos\ the $b$ and $c$ fields have conformal
dimensions one and zero respectively and they satisfy the OPE:
$$
b(z)\,c(w)\,=\,{1\over z-w}.
\eqn\ttresbis
$$

When trying to translate all expressions in terms of this $(b,c)$
system, one has to compute the
normal ordered products of $b$, $c$ and their
derivatives in terms of the field $\vec \phi$.
Some of these products that will be needed below are:
$$
\eqalign{
:bc:\,\, =&\,\,\vec\alpha\cdot\partial\vec\phi\cr
:b\partial c:\,\,=\,\,&{1\over 2}(\vec\alpha\cdot \partial^2\vec\phi-
(\vec\alpha\cdot\partial\vec\phi)^2 )\cr
:cb\partial c:\,\,=\,\,&{1\over 2}(\vec\alpha\cdot\partial^2\vec\phi
+(\vec\alpha\cdot\partial\vec\phi)^2 )
\,e^{-\vec \alpha\cdot\vec\phi}.\cr}
\eqn\tcuatro
$$
In \tcuatro\ by $:cb\partial c:$ we mean $:c(:b\partial c:):$. In order to
extract the contribution of the $(b,c)$ system to the generators of the  TA,
we shall decompose all vectors in components parallel and orthogonal to
$\vec \alpha$. Thus, if $\vec X$ is an arbitrary vector, we shall write :
$$
\vec X=\vec X_{\parallel}+\vec X_{\perp},
\eqn\tcinco
$$
where
$$
\eqalign{
\vec X_{\parallel}=&\,\, {(\vec X\cdot\vec\alpha)\over
\vec\alpha\,^{2}}\,\,\vec\alpha =\,\, -(\vec X\cdot\vec\alpha)\,\vec\alpha\cr
\vec X_{\perp}=\,\,&\vec X-X_{\parallel}=\,\,\vec X+
(\vec X\cdot\vec\alpha)\,\vec\alpha.\cr}
\eqn\tseis
$$
Using eqs. \vdos\ and \vocho\ one easily gets the components of $\vec A$,
$\vec R$ and $\vec g$ parallel to $\vec \alpha$:
$$
\vec A_{\parallel}=-{\vec \alpha\over 2}
\,\,\,\,\,\,\,\,\,\,\,\,\,\,\,\,\,
\vec R_{\parallel}=\vec\alpha
\,\,\,\,\,\,\,\,\,\,\,\,\,\,\,\,\,
\vec g_{\parallel}=-{d+1\over 2}\,\,\vec\alpha.
\eqn\tsiete
$$
Notice that the BRST charge $Q$ is simply the $b$ field :
$$
Q(z)=b(z).
\eqn\tocho
$$
To find the expression of $T$ and $R$ in the new variables, we first split
them as :
$$
T=T_{\parallel}+T_{\perp}
\,\,\,\,\,\,\,\,\,\,\,\,\,\,\,\,\,
R=R_{\parallel}+R_{\perp}.
\eqn\tnueve
$$
Making use of eq. \tsiete, the parallel components $T_{\parallel}$ and
$R_{\parallel}$ are easily obtained :
$$
\eqalign{
T_{\parallel}=&\,\, {1\over 2} (\vec\alpha\cdot\partial\vec\phi)^{2}
-{1\over 2}\vec\alpha\cdot\partial^2\vec\phi=\,\,-b\partial c\cr
R_{\parallel}=&\,\,\vec\alpha\cdot\partial\vec\phi=\,\,bc,\cr}
\eqn\cuarenta
$$
where we have taken into account the bosonization formulas of eq. \tcuatro.
Let us denote the components of $\vec \phi$, $\vec R$ and $\vec A$
orthogonal to $\vec \alpha$ as :
$$
\vec \varphi=\vec \phi_{\perp}
\,\,\,\,\,\,\,\,\,\,\,\,\,\,\,\,\,
\vec r=\vec R_{\perp}
\,\,\,\,\,\,\,\,\,\,\,\,\,\,\,\,\,
\vec a=\vec A_{\perp}.
\eqn\cuno
$$
Notice that $ \vec \varphi$, $\vec r$ and $\vec a$ live in an
$(N-1)$-dimensional vector space (\ie\ in the hyperplane orthogonal to
$\vec \alpha$). With these notations, $T$ and $R$ can be written as :
$$
\eqalign{
T=&\,\,T_{\varphi}+T_{(b,c)}=-{1\over 2}(\partial \vec \varphi)^2 +\vec
a\cdot\partial^2\vec\varphi -b\partial c\cr
R=&\,\, R_{\varphi}+bc=\vec r\cdot \partial \vec\varphi+bc.\cr}
\eqn\cdos
$$
It remains to express $G$ in terms of $b$, $c$ and $\vec\varphi$. This is
easily accomplished if one takes into account that the orthogonal
decomposition of the terms $\vec g\cdot\partial^2\vec \phi$ and
$\partial\vec\phi\cdot H\cdot\partial\vec\phi$ appearing in $G$ is:
$$
\eqalign{
\vec g\cdot\partial^2\vec \phi=&\,\, \vec a\cdot\partial^2\vec \varphi-
{d+1\over 2}\,\,\vec\alpha\cdot\partial^2\vec \phi\cr
\partial\vec\phi\cdot H\cdot\partial\vec\phi=
&-{1\over 2}(\partial\vec
\varphi)^2 +{d-1\over 2}\,\,(\vec\alpha\cdot\partial\vec\phi)^2-
R_{\varphi}\,\,\vec\alpha\cdot\partial\vec\phi,\cr}
\eqn\ctres
$$
where eqs. \vsiete\ and \vcuatro\ have been used. We have also used to
obtain \ctres\ the fact that $H$ acting on any two vectors orthogonal to
$\vec \alpha$ equals the matrix $-{I\over 2}$ :
$$
\vec X_{\perp}\cdot H\cdot\vec Y_{\perp}=-{\vec X_{\perp}\cdot\vec
Y_{\perp}\over 2}.
\eqn\ccuatro
$$
Eq. \ccuatro\ follows easily from eq. \vcinco. Substituting eq. \ctres\
in the second equation of \vtres\ and using the bosonization dictionary
of eq. \tcuatro, we obtain:
$$
G=\,\,cT_{\varphi}-cb\partial c +R_{\varphi}\partial c +{d\over
2}\partial^2  c,
\eqn\ccinco
$$
where $T_{\varphi}$ and $R_{\varphi}$ are defined in eq. \cdos. In terms
of $\vec \varphi$ and $(b,c)$ the realization of the TA of eqs. \tocho,
\cdos\ and \ccinco\ depends on two $(N-1)$-dimensional vectors $\vec a$
and $\vec r$. The constraints needed to ensure the closure of the TA can
be easily rephrased in terms of the scalar products of the vectors $\vec
a$ and $\vec r$. Putting $n=N-1$, we obtain from eqs. \cinco, \vseis\ and
\tsiete :
$$
\vec a\,^{2}={2-n\over 12}
\,\,\,\,\,\,\,\,\,\,\,\,\,\,\,\,\,
\vec r\,^{2}=1-d
\,\,\,\,\,\,\,\,\,\,\,\,\,\,\,\,\,
\vec a \cdot\vec r=-{1+d\over 2}.
\eqn\cseis
$$
For $n\geq 1$ it is possible to solve the constraints of eq. \cseis. In
fact when $n=1$ the value of $d$ is fixed ( $d$ can only be $-2$ or
$-{1\over 3}$), whereas for $n > 1$ $d$ remains arbitrary.

Let us consider now the realization of the TA that  is obtained by
performing a mirror transformation. After a redefinition as in eq. \once,
the energy-momentum tensor in bosonic language takes the form

$$
T^{*}=-{1\over 2} (\partial\vec \phi)^2 +
(\vec A-\vec R)\cdot\partial^2\vec\phi.
\eqn\csiete
$$
Notice that the mirror transformation changes the vector of the
background charges from $\vec A$ to $\vec A-\vec R$. More illuminating is
however the expression of $T^{*}$ in terms of the $(b,c)$ fields.
Taking  the expressions of $T$ and $R$ in eq. \cdos\ into account, one gets

$$
T^{*}=T^{*}_{\varphi}+T^{*}_{(b,c)}=-{1\over 2} (\partial\vec \varphi)^2
+\vec a\,^{*}\cdot\partial^{2}\vec\varphi -2b\partial c +c\partial b,
\eqn\cocho
$$
where
$$
\vec a\, ^{*}=\vec a -\vec r.
\eqn\cnueve
$$
In $T^{*}$  the fields $b$ and $c$ have weights $2$ and $-1$
respectively. Therefore this means that $T^{*}$ can be regarded as the
energy-momentum tensor of a bosonic string, being $(b,c)$ the standard
reparametrization ghosts. Other generators of the TA can be equally
obtained for this mirror realization. For example $G^{*}$ and $R^{*}$ are
given by
$$
\eqalign{
G^{*}=&\,\,b\cr
R^{*}=&\,\,R^{*}_{\varphi}+R^*_{(b,c)}\,=\,\,\vec r\,^{*}\cdot
\partial\vec\varphi-bc,\cr}
\eqn\cincuenta
$$
with $\vec r\,^{*}=-\vec r$. The new BRST current is
$$
Q^{*}=c\,[T^{*}_{\varphi} +{1\over 2} T^{*}_{(b,c)}\,]
-\partial \,[c\,(\,R^{*}_{\varphi}\,+\,{1\over 2}\,R^*_{(b,c)}\,)]
+{d\over 2}\partial^2 c.
\eqn\ciuno
$$
The first term in eq. \ciuno\   coincides with the standard BRST
current in the bosonic string theory. The second
and third terms in this equation  are total
derivatives and therefore they do not contribute to the integrated BRST
charge. Therefore the topological symmetry we are dealing with in
this mirror realization is just an improved version of the BRST
symmetry of the bosonic string [\gato]. Notice that the improving term
in \ciuno\ depends on $d$ and on the direction of the abelian current
$R^{*}$ in the space of fields. This mirror realization is given in
terms of two vectors $\vec a\, ^{*}$ and $\vec r\,^{*}$, which
determine respectively the background charges in the energy-momentum
tensor and the $R$-charges of the bosonic fields. It is
straightforward to obtain from eq. \cseis\ the constraints that $\vec
a\, ^{*}$ and $\vec r\,^{*}$ must satisfy :
$$
(\vec a \,^{*})^2=\,{26-n\over 12}
\,\,\,\,\,\,\,\,\,\,\,\,\,\,\,\,\,
\vec a \,^{*}\cdot \vec r\,^{*}=\,{3-d\over 2}
\,\,\,\,\,\,\,\,\,\,\,\,\,\,\,\,\,
(\vec r\,^{*})^{2}=1-d.
\eqn\cidos
$$

It is important to point out that for $d\not= 0$ the mirror transformation
can be implemented as a transformation of the fields. Let us come back to
the formalism in which only the bosonic fields appear (\ie\ before the
introduction of the $(b,c)$ pair). Consider the hyperplane orthogonal to
the vector $\vec R$. For any vector $\vec X$, a reflection with respect to
this hyperplane is a transformation that changes the sign of the component
of $\vec X$ parallel to $\vec R$. Calling   the reflected
vector $\vec X^{*}$, we have :
$$
\vec X\rightarrow \vec X\, ^{*}=\,\, \vec X-
{2(\vec X\cdot \vec R)\over \vec R\,^{2}}\,\, \vec R.
\eqn\citres
$$
As $\vec R\, ^{2}=-d$, eq. \citres\ reduces to :
$$
\vec X\, ^{*}=\,\, \vec X+ {2\over d}\,
(\vec X\cdot \vec R)\,\, \vec R,
\eqn\cicuatro
$$
which is a well-defined orthogonal transformation for $d\not= 0$. Let us
apply this transformation to the field $\vec \phi$, \ie\ let us define a
new field $\vec \phi\,^{*}$ as :
$$
\vec \phi\, ^{*}=\,\, \vec \phi+ {2\over d}\,
(\vec \phi\cdot \vec R)\,\, \vec R.
\eqn\cicinco
$$
In terms of $\vec \phi\,^{*}$, the energy-momentum tensor $T$ in \tres\
can be written as :
$$
T=-{1\over 2}(\partial \vec \phi\, ^{*})^2+
\vec A\,^{*}\cdot\partial^2\vec \phi^{*},
\eqn\ciseis
$$
where we have used the fact that the scalar product of two vectors is
invariant under orthogonal transformations. From the general formula
\cicuatro\ and the scalar product $\vec A\cdot\vec R$ given in eq. \vseis,
we get
$$
\vec A\,^{*}= \vec A-\vec R,
\eqn\cisiete
$$
which means that $T$ is given by :
$$
T=-{1\over 2}(\partial \vec \phi\, ^{*})^2+
(\vec A-\vec R)\cdot\partial^2\vec \phi^{*}.
\eqn\ciocho
$$
By comparing eqs. \ciocho\ and \csiete\ we conclude that the transformation
$T\rightarrow T-\partial R$ is equivalent to the change of variables
$\vec\phi\rightarrow \vec\phi\,^{*}$. Notice that with this interpretation
of the mirror transformation we do not change the energy-momentum tensor
(we just reexpress it in terms of new fields) and therefore the system and
its mirror copy can be regarded as the same model. As
$\vec R\, ^{*}=-\vec R$ (see eq. \citres), the $R$-current can be written
as :
$$
R=-\vec R\cdot\partial \vec \phi^{*},
\eqn\cinueve
$$
which means that the components of $R$ along the new field variables
$\vec \phi^{*}$ are minus the ones with respect to the initial fields
$\vec\phi$. Thus we see that changing  variables as
$\vec\phi\rightarrow \vec\phi\,^{*}$ without transforming $R$ is equivalent
to the transformation $R\rightarrow -R$ in eq. \once. In order to get the
full set of generators of the TA in terms of the $\vec\phi\,^{*}$ field,
let us consider the operators obtained by replacing $\vec\phi$ by
$\vec\phi\,^{*}$ in $Q$ and $G$. The conformal dimensions of the
vertex operators $e^{\pm \vec\alpha\cdot\vec\phi\,^{*}}$ with respect
to $T$ are : $$
\Delta ( e^{\vec \alpha \cdot\vec \phi\,^{*}})=\,\,2
\,\,\,\,\,\,\,\,\,\,\,\,\,\,\,\,\,
\Delta ( e^{-\vec \alpha \cdot\vec \phi\,^{*}})=\,\,-1,
\eqn\sesenta
$$
as  can be easily proved from eqs. \ciocho, \vdos\ and \vseis. It is
thus natural to consider the operators
$$
\eqalign{
Q^{*}=&(\vec g\cdot \partial ^2\vec \phi\,^{*}
+\partial\vec\phi\,^{*}\cdot H\cdot\partial\vec\phi\,^{*})
e^{-\vec \alpha\cdot\vec \phi\,^{*}}\cr
G^{*}=&\,e^{\vec \alpha\cdot\vec \phi\,^{*}}.\cr}
\eqn\suno
$$
It is easy to convince  oneself that $T$, $R$, $Q^{*}$ and $G^{*}$ close the
TA. Notice that the new BRST current $Q^{*}$ is formally obtained by
replacing $\vec \phi$ by $\vec\phi\,^{*}$ in $G$. However $Q^{*}\not= G$,
as  can be seen by writing $Q^{*}$ in terms of $\vec \phi$. Taking into
account that $\vec A\cdot\vec B\,^{*}=\vec A\,^{*}\cdot\vec B$ for any two
vectors $\vec A$ and $\vec B$, we get
$$
Q^{*}=(\vec g\,^{*}\cdot \partial ^2\vec \phi
+\partial\vec\phi\cdot H\cdot\partial\vec\phi)
e^{-\vec \alpha\,^{*}\cdot\vec \phi},
\eqn\sdos
$$
where we have taken into account eq. \cicinco\ and the fact that
$H\cdot\vec R=0$
(see eq. \tdos). On the  other hand from the general equation
\cicuatro\ and from eqs. \treinta\ and \tdos\ we have
$$
\vec g\,^{*}=\vec g
\,\,\,\,\,\,\,\,\,\,\,\,\,\,\,\,\,
\vec \alpha \,^{*}=\vec \alpha - {2\over d}\vec R.
\eqn\stres
$$
This implies that :
$$
Q^{*}=(\vec g\cdot \partial ^2\vec \phi
+\partial\vec\phi\cdot H\cdot\partial\vec\phi)
e^{-(\vec \alpha-{2\over d}\vec R)\cdot\vec \phi},
\eqn\scuatro
$$
and therefore $Q^{*}\not= G$ as stated above. In the same way $G^{*}$,
obtained by making the replacement of $\vec \phi$ by $\vec\phi\,^{*}$ in
$Q$,  differs from it and it is given by :
$$
G\,^{*}=e^{(\vec \alpha-\,\,{2\over d}\vec R)\cdot\vec \phi}.
\eqn\scinco
$$
It is easy to see how the reparametrization ghosts with conformal weights
$(2,-1)$ appear in the new variables. In complete analogy with eq.
\ttres, let us introduce two new fields $b^*$ and $c^*$ defined as
$$
b^{*}=e^{\vec \alpha \cdot\vec \phi\,^{*}}
\,\,\,\,\,\,\,\,\,\,\,\,\,\,\,\,\,\,\,
c^{*}=e^{-\vec \alpha \cdot\vec \phi\,^{*}}.
\eqn\sseis
$$
{}From eq. \sesenta\ it follows that $(b^*,c^*)$ have indeed conformal
dimensions $(2,-1)$ and thus we expect them to correspond to the
reparametrization ghosts that are needed to interpret our system as a
string theory. To verify this fact we proceed as we did previously with the
$(b,c)$ fields. Let us separate everywhere the component of $\vec\phi\,
^{*}$ parallel to $\vec \alpha$ from those that are orthogonal. If we
denote the latter by $\vec\varphi\,^{*}$, \ie\ if we define
$$
\vec \varphi \,^{*}=\vec \phi_\perp ^{*},
\eqn\ssiete
$$
it is straightforward to write down the expression of $T$ and $R$ in terms
of $\vec \varphi\,^*$, $b^*$, and $c^*$ :
$$
\eqalign{
T=&T_{\varphi^{*}}+T_{(b^{*},c^{*})}=
-{1\over 2} (\partial\vec \varphi\,^{*})^2
+\vec a\,^{*}\cdot\partial^{2}\vec\varphi\,^{*} -
2b^{*}\partial c^{*} +c^{*}\partial b^{*}\cr
R=&\,\,R_{\varphi^{*}}+R_{(b^{*},c^{*})}=\,\,\vec r\,^{*}\cdot
\partial\vec\varphi\,^{*}-b^{*}c^{*},\cr}
\eqn\socho
$$
where the vectors $\vec a\,^{*}$ and $\vec r\,^{*}$ are the same that
appear in eqs. \cocho\ and \cincuenta, and therefore they satisfy the
constraints of eq. \cidos. In the same way one can readily prove that
$Q^*$ and $G^*$ as given in eq. \suno\ can be written as
$$
\eqalign{
Q^{*}=&c^{*}\,[T_{\varphi^{*}} +{1\over 2} T_{(b^{*},c^{*})}]
-\partial \,[c^{*}\,(\, R_{\varphi^{*}}\,+\,{1\over 2}
\,R_{(b^{*},c^{*})}\,)]
+{d\over 2}\partial^2 c^{*}\cr
G^{*}=&\,b^{*}.\cr}
\eqn\snueve
$$
The comparison of eqs. \socho\ and \snueve\ with eqs. \cocho, \cincuenta\
and \cuno\ shows that, as expected, the change of variables
$\vec\varphi\rightarrow\vec\varphi\,^*$, $(b,c)\rightarrow (b^*,c^*)$ is
equivalent to the mirror redefinition of $T$, $R$, $Q$ and $G$. This fact
will be exploited in the next section, where we study the implications of
the topological symmetry in critical and non-critical bosonic string theory.

\chapter{The topological symmetry of the bosonic string theory}

In this section we analyse the string theory aspects of the realization of
the TA found in the previous section. In order to be coherent with the
notations used so far, all quantities related to the string theory
realization of the TA will be labelled by an asterisk, whereas those
variables and vectors obtained after a mirror redefinition of the string
fields will not carry such a label. First of all,
let us suppose that we have
a critical string theory. This means that the number $n$ of components of the
field $\vec \varphi\,^*$ is $26$ and that the background charge of
this field is zero (\ie\ that $\vec a\,^*=0$). As in this case
$\vec a\,^*\cdot\vec r\,^*$ obviously vanishes, from the constraints
generated by the TA (eq.\cidos) one gets that the value of the dimension
$d$ is
$$
d\,\,({\rm critical\,\,\, string})=3.
\eqn\setenta
$$
Notice that $d=3$ is obtained no matter the choice of $\vec r\,^*$. Of
course, the value of $(\vec r\,^*)^2$ is fixed by eq. \cidos. Taking
 eq. \setenta\ into account, we see that $(\vec r\,^*)^2=-2$,
which in particular means that $\vec r\,^*$ cannot be zero.

Suppose now that our string is a non-critical string, \ie\ that we identify
one of the $n$ components of $\vec \varphi\,^*$ with the Liouville field.
This identification can be achieved if we introduce an unitary vector
$\vec u_L$ such that its direction corresponds to the Liouville degree of
freedom, whereas the components of $\vec\varphi\,^{*}$ in the orthogonal
complement of $\vec u_L$ in the $n$-dimensional space represent the matter
fields. Accordingly, we decompose the vectors $\vec a\,^{*}$, $\vec
r\,^{*}$ and the field $\vec \varphi\,^{*}$ as

$$
\vec a\,^{*}=\,\vec a\,^{*}_{m}+\vec a\,^{*}_{L}
\,\,\,\,\,\,\,\,\,\,\,\,\,\,\,\,\,\,
\vec r\,^{*}=\,\vec r\,^{*}_{m}+\vec r\,^{*}_{L}
\,\,\,\,\,\,\,\,\,\,\,\,\,\,\,\,\,\,
\vec \varphi\,^{*}=\,\vec \varphi\,^{*}_{m}+\vec \varphi\,^{*}_{L}.
\eqn\stuno
$$
Expressed in  these new variables, the energy-momentum tensor $T$ can be
written as
$$
T=
-{1\over 2} (\partial\vec \varphi\,^{*}_m)^2
+\vec a\,^{*}_m\cdot\partial^{2}\vec\varphi\,^{*}_m
-{1\over 2} (\partial \varphi\,^{*}_L)^2
+ a\,^{*}_L\,\partial^{2}\varphi\,^{*}_L
-2b^{*}\partial c^{*} +c^{*}\partial b^{*},
\eqn\stdos
$$
where we have taken into account that $\vec a\,^{*}_L$ and
$\vec \varphi\,^{*}_L$ live in a one-dimensional vector subspace.
The central charge of the matter
can be obtained as a function of  $\vec a\,^{*}_m$. Indeed from eq. \stdos,
one has
$$
c^{*}_m=n-1+12 (\vec a\,^{*}_{m})^2.
\eqn\sttres
$$
Of course the Liouville field $\vec \varphi\,^{*}_L$ has central charge
$c^{*}_L=26-c^{*}_m$. As we have mentioned in the previous section, when
one analyzes the topological symmetry of the non-critical string, it is
convenient to introduce the field variables obtained after a mirror
redefinition. If we decompose the mirror field $\vec \varphi$ into its
components along the matter and Liouville directions (denoted by
$\vec \varphi_m$ and $\vec \varphi_L$ respectively), we can write the
energy-momentum tensor $T$ of the string given in eq. \stdos\ as:
$$
T=
-{1\over 2} (\partial\vec \varphi_m)^2
+\vec a_m\cdot\partial^{2}\vec\varphi_m
-{1\over 2} (\partial \varphi_L)^2
+ a_L\,\partial^{2}\varphi_L
-b\partial c,
\eqn\stcuatro
$$
where $(b,c)$ are $(1,0)$ ghosts obtained from the string fields as in eq.
\sseis\ (see below). In complete analogy with eq. \sttres, the central
charge of the mirror matter fields is given by:
$$
c_m=n-1+12(\vec a _m)^2.
\eqn\stcinco
$$
The precise relation between the string variables and their mirror
counterparts depends on the vector $\vec r \,^*$, which parametrizes the
different topological symmetries of the string theory. Recall (see eq.
\socho) that the components of $\vec r\,^*$ determine the topological
$U(1)$-charges of the field $\vec\varphi\,^*$, \ie\ the topological quantum
numbers of $\vec\varphi\,^*_m$ and $\vec\varphi\,^*_L$. In fact, the
projection of $\vec r\,^*$ into the matter hyperplane (denoted by
$\vec r\,^*_m$ in eq. \stuno) enters in the relation between
$c^{*}_m$ and $c_m$. Taking into account that
$\vec a_m=\vec a\,^*_m -\vec r\,^*_m$ in eq. \stcinco,
one immediately gets
$$
c_m=c^{*}_m+12(\vec r\,^{*}_{m})\,^2-
24 \vec a\,^{*}_m\cdot\vec r\,^{*}_{m}.
\eqn\stseis
$$
Let us now extract some information from the constraints \cidos\ generated
by the TA. Splitting $\vec a\,^*$ and $\vec r\,^*$  into their matter and
Liouville parts, the last two equations in \cidos\ can be written as :
$$
(\vec r\,^{*}_{m})\,^2+( r\,^{*}_{L})\,^2=1-d
\,\,\,\,\,\,\,\,\,\,\,\,\,\,\,\,\,\,\,\,
\vec a\,^{*}_m\cdot\vec r\,^{*}_{m}+ a\,^{*}_L r\,^{*}_{L}\,\,=
\,\,{3-d\over 2}.
\eqn\stsiete
$$
Eliminating $r\,^*_L$ in these equations, we get :
$$
\vec a\,^{*}_m\cdot\vec r\,^{*}_{m}+a\,^{*}_L
\,\sqrt{1-d-(\vec r\,^{*}_{m})\,^2}\,\,=\,{3-d\over 2},
\eqn\stocho
$$
which can be converted in a quadratic equation in the variable $d$. By
solving this equation, we obtain after a short calculation
$$
d\,\,=\,\,{{c_m+c^{*}_m\over 2}-\,7\,-\,6\,(\vec r\,^{*}_{m})\,^2
\, \pm\, \sqrt{(1-c_m)(25-c^{*}_m)}\over 6}.
\eqn\stnueve
$$
Eq. \stnueve\ gives the central extension $d$ of the TA as a function of the
matter central charges $c_m$ and $c^{*}_m$ and of the vector
$\vec r\,^{*}_{m}$ of the topological $U(1)$
 charges of the matter fields. Notice the remarkable fact that the term
under the square root in eq. \stnueve\ depends on the differences
$1-c_m$ and $25-c^{*}_m$. This means that the implementation of the
topological symmetry in these matter-plus-gravity models is determined
by the values of the matter central charge $c^{*}_m$ and of its mirror
counterpart. As a matter of fact, one can regard eq. \stnueve\ as a
``barrier" formula. Indeed, if we choose $(\vec r\,^{*}_{m})^2$ to be
real, the domains of values of  $c_m$ and $c^{*}_m$ for which $d$ is also
real are:
$$
c_m\leq 1 \,\,\,\,\,\,\,\, c^{*}_m\leq 25
\,\,\,\,\,\,\,\,\,\,\,\,\,\,\,
{\rm and}
\,\,\,\,\,\,\,\,\,\,\,\,\,\,\,
c_m\geq 1 \,\,\,\,\,\,\,\, c^{*}_m\geq 25   .
\eqn\ochenta
$$
Notice that  the central charges $c_m$ and $c^{*}_m$ appearing in
\ochenta\ are related by \stseis. Actually if the matter degrees of
freedom are neutral under the topological symmetry (\ie\ if $\vec
r\,^{*}_{m}=0$), $c_m$ and $c^{*}_m$ are equal
and \stnueve\ reduces to [\gato]
$$
d=\,\,{c^{*}_m-7\,\pm\,\,\sqrt{(1-c^{*}_m)(25-c^{*}_m)}\over 6} .
\eqn\ouno
$$
Requiring in this case  $d$ to be real, we obtain the standard range for
the matter central charge of the  bosonic non-critical strings, namely
$\{ c^{*}_m\leq 1\} \cup \{c^{*}_m\geq 25\}$. Notice that, taking
the minus sign in eq. \ouno, $d=0$ corresponds to the trivial matter
system of ref.
\REF\dist{J. Distler \journal\np&B342(90)523.}[\dist] in which
$c^{*}_m=-2$.

Let us now study  the form of the mirror change of variables for the
non-critical string theory. First of all we must
 bosonize the ghost fields $(b,c)$ and
$(b^*,c^*)$. For that purpose we introduce two scalar fields $\chi$ and
$\chi^*$, whose relation with the ghost fields is given by the standard
bosonization formulas, \ie\ by :
$$
\eqalign{
b=&e^{\chi}
\,\,\,\,\,\,\,\,\,\,\,\,\,\,\,\,\,\,
c=e^{-\chi}\cr
b^{*}=&e^{\chi^{*}}
\,\,\,\,\,\,\,\,\,\,\,\,\,\,\,
c^{*}=e^{-\chi^{*}}.\cr}
\eqn\odos
$$
The fields $\chi$ and $\chi^*$ have the OPE's
$$
\chi (z)\,\,\chi (w)=\,\chi^{*} (z)\,\,\chi^{*} (w)=\,log (z-w) .
\eqn\otres
$$
The contributions of the fields $\chi$ and $\chi^*$ to the energy-momentum
tensor can be computed by using standard methods from the expressions of
$T_{(b,c)}$ and $T_{(b^{*},c^{*})}$. The result is:
$$
\eqalign{
T_{(b,c)}=& {1\over 2} (\partial \chi)^2-{1\over 2}\partial ^2\chi\cr
T_{(b^{*},c^{*})}=& {1\over 2} (\partial \chi^{*})^2-
{3\over 2}\partial ^2\chi^{*}.\cr}
\eqn\ocuatro
$$
The change of variables corresponding to the mirror transformation was
given in eq. \cicinco\ for the fields $\vec\phi$ and  $\vec\phi\,^*$. The
decomposition of these fields along the ghost, Liouville and matter
directions is :
$$
\vec \phi =\, \vec \varphi_{m}+\vec \varphi_{L}
-\chi\,\,\vec \alpha
\,\,\,\,\,\,\,\,\,\,\,\,\,\,\,\,\,\,\,\,
\vec \phi\,^{*} =\, \vec \varphi\,^{*}_{m}+\vec \varphi\,^{*}_{L}
-\chi^{*}\,\vec \alpha,
\eqn\ocinco
$$
where we have taken into account that $\chi$ ($\chi^*$) parametrizes the
component of $\vec\phi$ ($\vec\phi\,^*$) parallel to
$\vec \alpha$ (see eqs.
\ttres\ and \sseis). Substituting eq. \ocinco\
in eq. \cicinco\ and using
the fact that $\vec R\cdot\vec\alpha =-1$ (see eq. \treinta)
one immediately
obtains the relation of $\vec \varphi\,^{*}_{m}$,
$\varphi\,^{*}_{L}$ and
$\chi^*$ with their mirror partners :
$$
\eqalign{
\vec \varphi\,^{*}_{m}\,=&\,(1+{2\over d}\,\,
\vec r\,^{*}_{m}\otimes\vec r\,^{*}_{m}\,\,)\,\vec \varphi_{m}
+{2r\,^{*}_{L}\vec r\,^{*}_{m}\over d}\,\varphi _{L}
+{2\over d}\,\vec r\,^{*}_{m}\,\chi\cr
\varphi\,^{*}_{L}\,=&\, {2r\,^{*}_{L}\over d}\,\,
\vec r\,^{*}_{m}\cdot\vec \varphi_{m}
+(1+{2(r\,^{*}_{L})^2\over d}\,)\,\,\varphi _{L}
+{2r\,^{*}_{L}\over d}\,\,\chi\cr
\chi^{*}\,=& -{2\over d}\,\,\vec r\,^{*}_{m}\cdot\vec \varphi_{m}
 -{2\over d}\,\,r\,^{*}_{L}\,\,\varphi _{L}\,
+\,(1-{2\over d})\,\,\chi .\cr}
\eqn\oseis
$$
By a direct calculation using eq. \oseis\ one
can check that the energy-momentum tensor when
expressed in terms of the mirror fields takes the form
displayed in eq. \stcuatro.

Let us now restrict ourselves for the remainder of this section
to the case $n=2$, \ie\ when there is only one
matter field. The minimal models coupled to two-dimensional gravity and the
$c=1$ string are particular cases of these $n=2$ theories. The constraints
\cidos\ now take  the form :
$$
\eqalign{
( a\,^{*}_{m})^2+( a\,^{*}_{L})^2=&\,\,2\cr
a\,^{*}_{m}\,r\,^{*}_{m}+a\,^{*}_{L}\,r\,^{*}_{L}=\,\,&{3-d\over 2}\cr
( r\,^{*}_{m})^2+( r\,^{*}_{L})^2=&\,\,1-d,\cr}
\eqn\osiete
$$
where the background charges $a\,^{*}_{m}$ and $a\,^{*}_{L}$ are determined
by the matter central charge $c\,^{*}_{m}$:
$$
a\,^{*}_{m}=\,\,i\,\sqrt{{1-c\,^{*}_{m}\over 12}}
\,\,\,\,\,\,\,\,\,\,\,\,\,\,\,\,\,\,\,\,\,\,
a\,^{*}_{L}=\,\,\sqrt{{25-c\,^{*}_{m}\over 12}}.
\eqn\oocho
$$
It is easy in this case to find the general solution of the constraints
dictated by the TA. First of all, let us parametrize the matter and
Liouville background charges as:
$$
a\,^{*}_{m}={i\lambda\over 2}\,\,-{i\over \lambda}
\,\,\,\,\,\,\,\,\,\,\,\,\,\,\,\,\,\,\,\,\,\,
a\,^{*}_{L}={\lambda\over 2}\,\,+{1\over \lambda} ,
\eqn\onueve
$$
where $\lambda$ is a function of $c\,^{*}_{m}$ that can be obtained by
comparing eq. \onueve\ with eq. \oocho.
Indeed after a short calculation one
arrives at the result
$$
\lambda=\,\,\sqrt{{25-c\,^{*}_{m}\over 12}}\,\pm\,
\sqrt{{1-c\,^{*}_{m}\over 12}}.
\eqn\noventa
$$
 The quantity $\lambda$ is well-known in non-critical string
theory. Indeed, the values of $\lambda$ in \noventa\ are such that the
vertex operator $e^{\lambda \varphi^*_L}$ has conformal weight one and thus
it can be considered as the gravitational dressing of the unit operator.
Usually in the Liouville theory the minus sign in \noventa\ is chosen in
order to make contact with the classical limit. Here we will not specify
any particular choice for this sign. Using the
representation \onueve\ it is straightforward to find the values of
$r\,^{*}_{m}$ and $r\,^{*}_{L}$ that satisfy eq. \osiete
$$
r\,^{*}_{m}=\,\, {i\over 4}(1-d)\lambda -\,{i\over \lambda}
\,\,\,\,\,\,\,\,\,\,\,\,\,\,\,\,\,\,\,\,\,\,
r\,^{*}_{L}=\,\, {1-d\over 4}\,\lambda +\,{1\over \lambda},
\eqn\nuno
$$
As the background charges $a_m$ and $a_L$ of the mirror fields are given by
$a_m=a^*_m-r^*_m$ and $a_m=a^*_m-r^*_m$, we find
$$
a_m=\,{i\over 4}\,(1+d)\lambda
\,\,\,\,\,\,\,\,\,\,\,\,\,\,\,\,\,\,\,\,\,\,
a_L=\,{1\over 4}\,(1+d)\lambda,
\eqn\ndos
$$
from which the central charge $c_m$
of the mirror fields in the matter direction
can be easily computed. Indeed, as in this case $c_m=1+12(a_m)^2$, it
follows that
$$
c_m=1-{(d+1)^2\over 8}\,[\,\,13-c\,^{*}_{m}\,\pm\,
\sqrt{(1-c\,^{*}_{m})(25-c\,^{*}_{m})}\,\,].
\eqn\ntres
$$
The double sign in eq. \ntres\ corresponds to
the two possible choices in
\noventa. Notice that for $d= -1$ one has $c_m\,=\, 1$
independently of the value of $c_m^*$. In this case $a_m=a_L=0$ as
it can be verified by inspecting eq. \ndos. For $d\not= -1$ one can
invert eq. \ntres\ and obtain the
expression of $c\,^{*}_{m}$ in terms of $c_m$. One gets:
$$
c\,^{*}_{m}=13-4\,{1-c_m\over (d+1)^2}\,-9\,{(d+1)^2\over 1-c_m}.
\eqn\ncuatro
$$

The topological symmetry of the string theory we have been analyzing depends
on the parameter $d$. For each value of $d$ we have a BRST current
$Q^*(z)$. Although  $Q^*(z)$'s for different $d$'s are different, the
corresponding integrated BRST charges are equal due to the fact that $d$
only enters in the total derivative terms of $Q^*(z)$. Thus $d$
parametrizes the freedom one has to choose an improving term in the BRST
current of string theory compatible with the TA. As we have seen,
associated to each BRST current, we have a set of fields that naturally
implement the mirror version of the string topological symmetry. It could
be that, when the string theory is formulated in terms of the mirror
variables, new symmetries that are hidden or realized in a non-local way in
the original fields, become apparent. Let us see some examples of how
this can occur.

First of all, let us ask ourselves if there is any particular value
of $d$ for which the spin one ghosts $(b,c)$ can be written locally
in terms of the reparametrization ghosts $(b^*,c^*)$. A glance at
eq. \oseis\ reveals that this only occurs for $d=1$. Indeed using
eqs. \odos\ and \oseis\ we obtain in this case:
$$
b\,=\,e^{2\vec r\,^*\cdot\vec\varphi\,^*}\,c^*
\,\,\,\,\,\,\,\,\,\,\,\,\,\,\,\,\,\,
c\,=\,e^{-2\vec r\,^*\cdot\vec\varphi\,^*}\,b^*,
\eqn\exuno
$$

Spin one ghosts are characteristic of topological sigma models [\wit].
Thus it is natural to think that in the mirror variables one could
formulate string theory as a topological sigma model. Moreover $d=1$ is
the natural value for the c-number anomaly of the BRST symmetry of a
topological sigma model whose target space has complex dimension
one. However,  a difficulty arises when one tries to identify the
$(b,c)$ fields with the ghosts of the topological sigma model. The
problem is that, with respect to the $R$-current of the string, the
fields $b$ and $c$ have $R$-charge $+1$ and $-1$ respectively. These
values of the $R$-charge are  opposite to the ones expected for
the ghosts of a
topological sigma model, where the antighost must have
conformal weight one. In order to overcome this difficulty one must
redefine $b$ and $c$. A simple exchange does not work since $b$ and
$c$ have different conformal dimensions. Actually, one can redefine
$b$ and $c$ in such a way that their conformal weights are
interchanged and their $R$-charges are not altered. The key
observation to find out the form of this redefinition is the fact
that $(\vec r\,^*)^2\,=\,0$ for $d=1$ (see eq. \osiete). This means
that we can use the exponentials of $\vec r\,^*\cdot \vec
\varphi\,^*$ to dress the fields $(b,c)$ in such a way that their
$U(1)$ charges are unaffected. Let us define new ghost fields $B$
and $C$ as:
$$
B\,=\,e^{\vec r\,^*\cdot\vec\varphi\,^*}\,c
\,\,\,\,\,\,\,\,\,\,\,\,\,\,\,\,\,\,
C\,=\,e^{-\vec r\,^*\cdot\vec\varphi\,^*}\,b.
\eqn\exdos
$$
The conformal weights of $B$ and $C$ are
$\Delta (B)\,=\,1$ and $\Delta (C)\,=\,0$. In terms of the
diffeomorphism ghosts $(b^*,c^*)$ $B$ and $C$ are given by:
$$
B\,=\,e^{-\vec r\,^*\cdot\vec\varphi\,^*}\,b^*
\,\,\,\,\,\,\,\,\,\,\,\,\,\,\,\,\,\,
C\,=\,e^{\vec r\,^*\cdot\vec\varphi\,^*}\,c^*.
\eqn\extres
$$
Using the value of $\vec r \,^*$ for $d=1$ (see eq. \nuno), $B$ and
$C$ can be written as a function of the matter and Liouville fields
of string theory:
$$
B\,=\,e^{{i\over\lambda}\,(\varphi_m^*\,+\,
i \varphi_L^*\, ) }\,b^*
\,\,\,\,\,\,\,\,\,\,\,\,\,\,\,\,\,\,
C\,=\,e^{-{i\over\lambda}\,(\varphi_m^*\,+\,
i \varphi_L^*\, ) }\,c^*.
\eqn\excuatro
$$
Notice that for a minimal
$(p,q)$ model with
$c_m^*\,=\,1\,-\,6\,{(p-q)^2\over pq}$ and $q\geq p$,  $\lambda$ is
equal to $ \sqrt{{2p\over q}}$ ( we are taking the minus sign in eq.
\noventa).

In order to verify that we are mapping string theory to a
topological sigma model we need to introduce new fields that could
represent coordinates in a target manifold. With this purpose in
mind, let us define two new fields $x$ and $\bar p$ as follows:
$$
\eqalign{
x\,=&\,[\,b^*c^*\,-\,{i\lambda\over 2}\,
(\partial\varphi_m^*\,-\,i \partial\varphi_L^*\,)\,]\,
e^{{i\over\lambda}\,(\varphi_m^*\,+\,
i \varphi_L^*\, ) }\cr\cr
\bar p\,=&\,e^{-{i\over\lambda}\,(\varphi_m^*\,+\,
i \varphi_L^*\, ) }.\cr}
\eqn\excinco
$$
The energy-momentum tensor $T$ and the $U(1)$ current $R$ of eq.
\socho\ have the following expressions in terms of $x$ and $\bar p$:
$$
\eqalign{
T\,=&\,-\partial x\bar p\,-\,B\partial C\cr
R\,=&\,-BC.\cr}
\eqn\exseis
$$
After identifying $\bar p\equiv \partial \bar x$, where $\bar x$ is
the holomorphic component of the complex conjugate of the coordinate
$x$, the operator $T$ in eq. \exseis\ represents the energy-momentum
tensor of a topological sigma model in one complex dimension
[\wit]. The fields $B$, $C$, $x$ and $\bar p$ satisfy the OPE's:
$$
B(z)C(w)\,=\,x(z)\bar p(w)\,=\,{1\over z-w}.
\eqn\exsiete
$$
On the other hand, the operators $Q^*$ and $G^*$ of eq. \snueve\ for
$d=1$ can be written in the remarkably simple form:
$$
Q^*\,=-C\partial x
\,\,\,\,\,\,\,\,\,\,\,\,\,\,\,\,\,\,
G^*\,=\,B\bar p.
\eqn\exocho
$$
In the form of eq. \exocho\ $Q^*$ is the standard topological
current for a topological sigma model. We have thus succeeded in
finding a set of fields in terms of which the BRST symmetry of
non-critical string theories is equivalent to the topological
symmetry of a $d=1$ topological sigma model. For $c_m^*\,=\,1$ the
above mapping was obtained in
\REF\IK{H. Ishikawa and M. Kato\journal\ijmp&A(94)5796.}
\REF\LlatasRoy{P.M. Llatas and S. Roy\journal\pl&B345(95)6.}
ref. [\IK]. This analysis was extended
to $c_m^*\,<\,1$ string theories in ref. [\LlatasRoy]. In these papers
it is shown how the BRST cohomology of the non-critical string
theory can be recovered from the physical states of the topological
sigma model. Notice that the field $x$ is nothing but one of the ground
ring generators of $c_m^*\,\leq 1$ string theory
\REF\WittenGR{E. Witten\journal\np&B373(92)187.}
\REF\Kutasov{D. Kutasov, E. Martinec and N.
Seiberg\journal\pl&B276(92)437.}[\WittenGR, \Kutasov].

As a second example of a non-trivial mapping generated by the
mirror transformation,  let us suppose next that we choose in eqs.
\nuno\ and \ndos\  the particular value of $d$ given by
$$
d=\,-1-{2\over\lambda} .
\eqn\ncinco
$$
For this value of $d$, the background charges of the mirror fields are:
$$
a_m=\,-{i\over 2}
\,\,\,\,\,\,\,\,\,\,\,\,\,\,\,\,\,\,\,\,\,\,
a_L=\,-{1\over 2},
\eqn\nseis
$$
which imply that our mirror system has central charge $c_m=-2$ in the
matter sector. It turns out that when eq. \nseis\ holds, there exist three
dimension-one operators that close a $SL(2)$ current algebra. In terms of
the mirror fields these three currents are:
$$
\eqalign{
H=&i\partial\varphi_m+\partial\varphi_L\cr
J_+=&{i\partial \chi\over \sqrt{2}}\,\,e^{-\chi-\varphi_L}\cr
J_-=&\,\,[\sqrt{2}(\partial\varphi_m-i\partial\varphi_L)-\,
{ i\partial\chi\over \sqrt{2}}\,\,]\,\,e^{\chi+\varphi_L}.\cr}
\eqn\nsiete
$$
Using the expression of $T$ given in eqs. \stcuatro\ and \ocuatro, one can
easily check that $H$ and $J_{\pm}$ are primary operators with conformal
dimension one. Moreover they close a $SL(2)$ current algebra without
central extension:
$$
\eqalign{
H(z)\,J_{\pm}(w)=&\,\,\pm \,{J_{\pm}(w)\over z-w}\cr
J_+(z)\,J_-(w)=&\,\,{H(w)\over z-w}\cr
H(z)\,H(w)=&J_+(z)\,J_+(w)=J_-(z)\,J_-(w)=0.\cr}
\eqn\nocho
$$
Actually eq. \nsiete\ is nothing but the standard free field realization of
a zero level $SL(2)$ current algebra
\REF\gera{A. Gerasimov et al. \journal\ijmp&A5(90)2495.}[\gera].

The
energy-momentum tensor of the string theory can be written as a Sugawara
bilinear in $H$ and $J_{\pm}$:
$$
T=\,\,{1\over 2}\,\,[\,:J_+J_-:+:J_-J_+:+:H^2:\,\,],
\eqn\nnueve
$$
where the global coefficient ${1\over 2}$ is the value corresponding to a
$SL(2)$ current algebra of vanishing level. Although eqs. \nocho\ and
\nnueve\ are easy to check when the mirror fields are used, they are not so
evident when the currents are  given in terms of the original fields of
the string theory. As a matter of fact,
the expressions of $H$ and $J_{\pm}$ in terms
of $\varphi^*_m$, $\varphi^*_L$ and $\chi^*$ can be worked out if one uses
the values of $d$, $r^*_m$ and  $r^*_L$ given in eqs. \ncinco\ and \nuno.
The result is:
$$
\eqalign{
H=&\,{1\over \lambda(\lambda+2)}\,[\,i(4-\lambda^2)\,\,\partial
\varphi\,^{*}_{m}\,-\,(4+\lambda^2)\,\,\partial\varphi\,^{*}_L
\,+\,4\lambda\,\, \partial \chi^*\,]\cr
J_{\pm}=&\, {{\cal J}_{\pm}\over \sqrt{2}}\,\,e^{\pm \Lambda},\cr}
\eqn\cien
$$
where ${\cal J}_{\pm}$ and $\Lambda$ are given by :
$$
\eqalign{
{\cal J}_+=&\,\, (\lambda-1)\,\,\partial\varphi\,^{*}_{m}\,+\,
i(1-\lambda-{4\over \lambda+2})\,\partial\varphi\,^{*}_L\,\,+
i(1+{2\lambda\over \lambda+2}\,)\,\partial\chi^{*}\cr
{\cal J}_-=&\,\,(\,{4\over\lambda}-\lambda-1)\,\,\partial\varphi\,^{*}_{m}\,
+\,i(1+\lambda +{4\over\lambda}-{4\over
\lambda+2})\,\partial\varphi\,^{*}_L\, -\,i(3+{4\over
\lambda+2})\,\partial\chi^*\cr \Lambda=&\,\, {\lambda+2\over
2\lambda}\,\,
[i(\lambda^2-1)\varphi\,^{*}_{m}\,+(\lambda^2+1)\varphi\,^{*}_{L}\,-
2\lambda\chi^{*}].\cr}
\eqn\ctuno
$$
{}From eqs. \cien\ and \ctuno\ one can easily check that, for any
value of $\lambda$, the currents $J_{\pm}$ and $H$ satisfy the centerless
algebra of eq. \nocho. Moreover, using  eqs. \cien\ and \ctuno, it is
straightforward to prove that, indeed, the Sugawara expression for $T$
given in eq. \nnueve\ reduces to the standard form of the
energy-momentum tensor of the non-critical strings, where the
background charges of the matter and Liouville fields are given in eq.
\onueve. Notice that $H$ and $J_{\pm}$, as displayed in eqs. \cien\ and
\ctuno, are in general non-local in the string ghost fields $(b^*,c^*)$.
They are however local in the spin one ghosts $(b,c)$. In fact the
ghost $b$ (\ie\ the mirror BRST charge $Q$, see eq. \tocho) is one of
the screening operators of the $SL(2)$ current algebra of eq. \nocho.
It would be interesting to investigate
the consequences of the $SL(2)$  symmetry of the non-critical bosonic
string that we have just uncovered. We will not attempt this here.
Instead we shall generalize in the next section our results to the
case of superconformal matter.

\chapter{Topological superconformal matter and superstrings}

The purpose of this section is to extend the results obtained so far to
conformal topological matter systems that
possess a superconformal
symmetry. We shall realize this supersymmetry in
terms of a multicomponent scalar field $\vec \phi$, as in the previous
section, and an $N$-component
 Majorana fermion
$\vec \Psi(z)=(\Psi_1(z),\ldots,\Psi_N(z))$. The OPE's among the
components of $\vec \Psi$ are given by :

$$
\Psi_i(z)\,\,\Psi_j(w)=-{\delta_{ij}\over z-w}.
\eqn\susyuno
$$
The holomorphic component of the energy-momentum tensor $T$ in given by :
$$
T=-{1\over 2}\partial\vec\phi\cdot\partial\vec\phi+
\vec A\cdot\partial^2\vec\phi+
{1\over 2}\vec\Psi\cdot\partial\vec\Psi .
\eqn\susydos
$$
The vanishing of the conformal anomaly $c$ of the operator $T$ in eq.
\susydos\ yields the following condition for the modulus of the
background charge $\vec A$:
$$
\vec A\,^{2}=-{N\over 8} .
\eqn\susytres
$$
The superconformal algebra in this model is generated by a fermionic
dimension-${3\over 2}$ operator that we shall denote by $T_F$. The
primary character of $T_F$ fixes the OPE of $T$ with $T_F$:
$$
T(z)\,\,T_F(w)={3\over 2}\,\,{T_F(w)\over (z-w)^2} +{\partial
T_F(w)\over z-w}.
\eqn\susycuatro
$$
Moreover, the OPE of the supersymmetry generator $T_F$ with itself
contains a single pole singularity whose residue is proportional to the
energy-momentum tensor $T$:
$$
T_F(z)\,\,T_F(w)={1\over 2}\,\,{T(w)\over z-w}.
\eqn\susycinco
$$
Notice that, due to the vanishing of the central charge, there are no
higher order poles in the right-hand side of eq.\susycinco. In order to
represent $T_F$ in terms of the fields $\vec \phi$ and $\vec \Psi$, we
shall adopt the following ansatz:
$$
T_F\,\,=\,\, {1\over 2}\,\,\partial\vec\phi\,\cdot L\cdot\vec\Psi+
\vec M\cdot\partial\vec\Psi+
K_{lmn}\Psi_l\Psi_m\Psi_n ,
\eqn\susyseis
$$
where $\vec M$, $L$ and $K$ are numerical tensors of ranks one, two and
three respectively. In \susyseis\ we have included all possible terms
compatible with the dimension and statistics of $T_F$. By requiring the
fulfillment of eqs. \susycuatro\ and \susycinco\ one arrives at the
following conditions on $\vec M$, $L$ and $K$:
$$
\eqalign{
L\,\,L^T=&\,\,L^T\,\,L=\,\,1\cr
\vec M=&-L\cdot\vec A\cr
K_{lmn}=&0.\cr}
\eqn\susysiete
$$
Substituting the values given by eq. \susysiete\ in eq. \susyseis, one
gets $T_F$ in terms of the orthogonal matrix $L$:
$$
T_F=\,\,{1\over 2}\,\, \partial\vec \phi\cdot L\cdot\vec \Psi
-\vec A\,\cdot L\cdot\partial\vec\Psi .
\eqn\susyocho
$$
Actually we can absorb the unknown matrix $L$ by redefining the fermionic
field $\vec \Psi$ as $\vec \Psi\rightarrow L\vec\Psi$. After this
redefinition $T_F$ takes the form:
$$
T_F=\,\,{1\over 2} \,\,\partial\vec \phi\cdot\vec \Psi
-\vec A\cdot\partial\vec\Psi.
\eqn\susynueve
$$
The presence of a superconformal symmetry in the chiral algebra of our
system suggests that all fields of the model can be arranged in
supersymmetry doublets. In general one of those doublets
$(X,Y)$ is composed by two primary fields $X$ and $Y$ whose conformal
weights $\Delta_X$ and $\Delta_Y$ differ by ${1\over2}$
($\Delta_Y=\Delta_X\,\, +\,\,{1\over 2}$). The action of $T_F$ on $X$
and $Y$ is given by:
$$
\eqalign{
T_F(z)\,\,X(w)=&\,\,{1\over 2}\,\, {Y(w)\over z-w}\cr\cr
T_F(z)\,\,Y(w)=&\,\,{\Delta_X X(w)\over (z-w)^2}+
{1\over 2}\,\, {\partial X(w)\over z-w}.\cr}
\eqn\susydiez
$$

Let us now consider the algebra closed by $T_F$ and the generators of
the topological symmetry. In the type of
models we are dealing with, it is natural to think that  the operators
$T$, $G$, $Q$ and $R$ will have a supersymmetric partner. As  can be
checked by comparing eqs. \susycinco\ and \susycuatro\ with the general
equation \susydiez, $T_F$ is the partner of $T$. Let us denote by $Q_B$,
$G_B$ and $R_F$  the partners of $Q$, $G$ and $R$ respectively. With
these fields we form four supersymmetry doublets
$(T_F,T)$, $(Q_B,Q)$, $(G_B,G)$ and $(R_F,R)$. The conformal weights of
the lower components of these doublets are:
$$
\eqalign{
\Delta (T_F)=&\,\,{3\over 2}
\,\,\,\,\,\,\,\,\,\,\,\,\,\,\,\,\,\,
\Delta (G_B)=\,\,{3\over 2}\cr\cr
\Delta (Q_B)=&\,\,{1\over 2}
\,\,\,\,\,\,\,\,\,\,\,\,\,\,\,\,\,\,
\Delta (R_F)=\,\,{1\over 2}.\cr}
\eqn\susyonce
$$

An algebra with the field content described above can be obtained by
twisting the $N=3$ superconformal algebra [\yos]. Let us describe this
algebra with our notations. First of all, $T$, $G$, $Q$ and $R$ close
the same algebra as in the non-supersymmetric case. On the other
hand, the new fields $T_F$, $G_B$, $Q_B$ and $R_F$ are primary with
respect to $T$ with the conformal weights displayed in eq. \susyonce.
The action of $T_F$ on the generators is determined by the
arrangement of these fields in supersymmetry doublets described above
(see eq. \susydiez). There is however an exception. The action of
$T_F$ on $(R_F,R)$ is anomalous. Actually one has:
$$
\eqalign{
T_F(z)\,\,R_F(w)=&\,\,-\,\,{{d\over 2}\over (z-w)^2}+
{1\over 2}\,\, {R(w)\over z-w}\cr\cr
T_F(z)\,\,R(w)=&\,{1\over 2}\,\,{R_F(w)\over (z-w)^2}+
{1\over 2}\,\, {\partial R_F(w)\over z-w}.\cr}
\eqn\susydoce
$$
The anomaly term in eq. \susydoce\ is related to the anomalous behaviour
of $R$ under the action of $T$. Two of the multiplets ($(T_F,T)$ and
$(Q_B,Q)$) are connected to the other two($(G_B,G)$ and
$(R_F,R)$) by the action of the BRST charge $Q$: the two latter
doublets are the BRST ancestors of the former ones. In fact, the
supersymmetry generator $T_F$ is $Q$-exact. Its $Q$-ancestor is $G_B$
and when one tries to relate  $G_B$ and $T_F$ by the action of $Q$,
one gets two new fields ($R_F$ and $Q_B$):
$$
\eqalign{
Q(z)\,\,G_B(w)=&-{1\over 2}\,\, {R_F(w)\over (z-w)^2}-
{T_F(w)\over z-w}\cr\cr
Q(z)\,\,T_F(w)=&-{1\over 2}\,\,\, {Q_B(w)\over (z-w)^2}.\cr}
\eqn\susytrece
$$
Notice that $R_F$ and $R$ play the same role with respect to $T_F$
and $T$ respectively (compare eq. \susytrece\ with eq. \nueve). These
two new fields $R_F$  and $Q_B$ are connected by $Q$:
$$
\eqalign{
Q(z)\,\,R_F(w)=&{Q_B(w)\over z-w}\cr\cr
Q(z)\,\,Q_B(w)=&0.\cr}
\eqn\susycatorce
$$
The non-vanishing singularities of the products of $Q_B$, $G_B$ and
$R_F$ with themselves are given by:
$$
\eqalign{
Q_B(z)\,\,G_B(w)=&\,\,-\,\,{{d\over 2}\over (z-w)^2}-
{1\over 2}\,\, {R(w)\over z-w}\cr\cr
R_F(z)\,\,R_F(w)=&\,\,\, {d\over z-w}.\cr}
\eqn\susyquince
$$
Finally, in order to complete the non-zero OPE's of the algebra, let
us display those of $G$ with the new generators:
$$
\eqalign{
G(z)\,\,T_F(w)=&-{3\over 2}\,\, {G_B(w)\over (z-w)^2}-
{\partial G_B(w)\over z-w}\cr\cr
G(z)\,\,Q_B(w)=&{1\over 2}\,\, {R_F(w)\over (z-w)^2}-
{ T_F(w)-\partial R_F(w)\over z-w}\cr\cr
G(z)\,\,G_B(w)=&0\cr\cr
G(z)\,\,R_F(w)=&-{G_B(w)\over z-w},\cr}
\eqn\susydseis
$$
and those of $R$ with $Q_B$, $G_B$ and $R_F$:
$$
\eqalign{
R(z)\,\,Q_B(w)=&\,\, {Q_B(w)\over z-w}\cr\cr
R(z)\,\,G_B(w)=&\,\, -{G_B(w)\over z-w}\cr\cr
R(z)\,\,R_F(w)=&0.\cr}
\eqn\susydsiete
$$
{}From eqs. \susyquince\ and \susydsiete\ one notices that the
operators $Q_B$, $G_B$ and $R$ close an $SL(2)$ current algebra with
level $k=2d$, which is a remnant of the $SL(2)$ Kac-Moody subalgebra
present in the $N=3$ superconformal algebra (before performing the
topological twist $Q_B$, $G_B$ and $R$ have conformal dimension
one). In what follows the algebra just described will be called the
supersymmetric topological algebra (STA). Notice that the STA is
nothing but the BRST algebra of the $N=1$ superconformal symmetry.

As  happened in the non-supersymmetric case, the STA enjoys a
mirror symmetry whose origin comes from the two possible
topological twists of the $N=3$ superconformal algebra. Under this
symmetry, from a given realization of the STA, we generate a new one
by changing $T$, $Q$ , $G$ and $R$ into  $T^*$, $Q^*$ , $G^*$ and $R^*$
as in eq. \once. The other four generators of the STA must be changed
as follows:  $$
\eqalign{
T_F\rightarrow T_F^{*}=&\,\,T_F-\partial R_F\cr
Q_B\rightarrow Q_B^{*}=&\,\,G_B\cr
G_B\rightarrow G_B^{*}=&\,\,Q_B\cr
R_F\rightarrow R_F^{*}=&-R_F.\cr}
\eqn\susydocho
$$
This mirror symmetry will play an important role in our analysis,
specially in connection with superstring theories.

We are now going to consider the realization of the STA in our
supersymmetric field theory. We could proceed as in section 2 and
study the possible vertex operator representations for the fields
appearing in the STA. Instead of doing that, we shall follow a
shorter path, which  takes into account the lessons learnt in the
analysis of section 2. As a direct generalization of eq. \tocho, we
represent the doublet $(Q_B,Q)$ by a doublet of fields $(\beta, b)$ :

$$
Q_B=\beta
\,\,\,\,\,\,\,\,\,\,\,\,\,\,\,\,\,\,
Q=b,
\eqn\susydnueve
$$
where $\beta$ ($b$) is a commuting (anticommuting) field. From eq.
\susyonce\ we see that the conformal weights of $\beta$ and $b$ must
be $\Delta (\beta)=\,{1\over 2}$ and $\Delta (b)=\,1$ respectively.
At this point  the question arises  of how one can represent $\beta$
and $b$ in terms of the fields $\vec \phi$ and $\vec \Psi$. Following
the spirit of the manifest supersymmetric bosonization of refs.
\REF\Martinec{E. Martinec and G. Sotkov\journal\pl&B208(88)240.}
\REF\Takama{M. Takama\journal\pl&B210(88)153.}
 [\Martinec, \Takama],
let us write:
$$
\eqalign{
\beta\,=&\,e^{\vec \mu\cdot \vec\phi}\cr\cr
b\,=&\,-\vec\mu\cdot\vec\Psi\,\,e^{\vec \mu\cdot \vec\phi},\cr}
\eqn\susyveinte
$$
where $\vec\mu$ is a numerical vector. It can be easily checked
that, indeed, the fields $(\beta, b)$ in eq. \susyveinte\ form a
supersymmetry doublet. Together with $(\beta, b)$ we must introduce
a conjugate doublet $(c,\gamma)$, where $c$ ($\gamma$) is an
anticommuting (commuting) field of dimension $0$ (${1\over 2}$)
which is conjugate to $b$($\beta$). From now on we will refer to
$b$, $c$, $\beta$ and $\gamma$ as the ghost fields. These fields can
be represented as:
$$
\eqalign{
c\,=&\, -\vec\lambda\cdot\vec\Psi\,\,
e^{-\vec \mu\cdot\vec\phi}\cr\cr
\gamma\,=&\,[\,\vec\lambda\cdot\partial\vec\phi\,-\,
(\vec\mu\cdot\vec
\Psi)\,(\vec\lambda\cdot\vec
\Psi)]\,\, e^{-\vec \mu\cdot\vec\phi},\cr}
\eqn\susyvuno
$$
where $\vec \lambda$ is a new numerical vector. In fact if we
require that $(c,\gamma)$ satisfy the first equation in \susydiez,
we obtain the following condition that must be satisfied by
$\vec\lambda$:
$$
\vec A\cdot\vec\lambda\,=\,{1\over 2},
\eqn\susyvdos
$$
which can be obtained by imposing the absence of the double pole
singularity in the product $T_F(z)c(w)$. The fact that the doublets
$(\beta, b)$ and $(c,\gamma)$ are conjugate is reflected in the
fact that the non-vanishing OPE's among these fields are:
$$
b(z)\,c(w)\,=\,{1\over z-w}
\,\,\,\,\,\,\,\,\,\,\,\,\,\,\,\,
\beta(z)\,\gamma(w)\,=\,{-1\over z-w}.
\eqn\susyvtres
$$
After a simple calculation, one can verify that the two ghost
doublets are conjugates if the scalar products of $\vec \mu$ and
$\vec \lambda$ are given by:
$$
\vec\lambda\cdot\vec\lambda\,=\,\vec\mu\cdot\vec\mu\,=\,0
\,\,\,\,\,\,\,\,\,\,\,\,\,\,\,\,\,\,\,\,\,
\vec\lambda\cdot\vec\mu\,=\,-1.
\eqn\susyvcuatro
$$
On the other hand, $\beta$ has conformal weight ${1\over 2}$ if $\vec
A\cdot\vec\mu$ is given by:
$$
\vec A\cdot \vec\mu\,=\,{1\over 2}.
\eqn\susyvcinco
$$

In order to separate the contributions of the ghost fields from the
remaining degrees of freedom of our system, it is convenient to
introduce two vectors $\vec \alpha_1$ and $\vec \alpha_2$, which are
linearly related to $\vec \mu$ and $\vec \lambda$ as:
$$
\vec \alpha_1\,=\,{1\over \sqrt 2}\, (\vec \mu+\vec \lambda)
\,\,\,\,\,\,\,\,\,\,\,\,\,\,\,\,\,\,\,\,\,\,
\vec \alpha_2\,=\,{i\over \sqrt 2}\, (\vec \mu-\vec \lambda),
\eqn\susyvseis
$$
where the imaginary unit $i$ has been included in the definition of
$\vec\alpha_2$ for convenience. These vectors satisfy:
$$
\vec \alpha_i\cdot\vec\alpha_j\,=\,-\delta_{ij}.
\eqn\susyvsiete
$$
It is now straightforward to split any vector $\vec X$ in its
components parallel and orthogonal to the ghost plane:
$$
\vec X=\vec X_{\parallel}+\vec X_{\perp},
\eqn\susyvsiete
$$
where
$$
\vec X_{\parallel}=\,\, {(\vec X\cdot\vec\alpha_1)\over
\vec\alpha_1\,^{2}}\,\,\vec\alpha_1 +
{(\vec X\cdot\vec\alpha_2)\over
\vec\alpha_2\,^{2}}\,\,\vec\alpha_2=\,\,
-(\vec X\cdot\vec\alpha_1)\,\vec\alpha_1
-(\vec X\cdot\vec\alpha_2)\,\vec\alpha_2.
\eqn\susyvnueve
$$
Using this last equation, together with the values of
$\vec A\cdot\vec\lambda$ and $\vec A\cdot\vec\mu$ given in eqs.
\susyvdos\ and \susyvcinco, it is straightforward to obtain the
component of the background charge $\vec A$ parallel to the ghost
plane:
$$
\vec A_{\parallel}=-{\vec \alpha_1\over \sqrt 2}=
-{1\over 2}\,(\vec \mu+\vec\lambda).
\eqn\susytreinta
$$
{}From this result we can split $T$ and $T_F$ as:
$$
T=T_{\parallel}+T_{\perp}
\,\,\,\,\,\,\,\,\,\,\,\,\,\,\,\,\,
T_F=T_F^{\parallel}+T_F^{\perp},
\eqn\susytuno
$$
with
$$
\eqalign{
T_{\parallel}\,\,=&\,\,
{1\over 2} (\vec\alpha_1\cdot\partial\vec\phi)^{2}+
{1\over 2} (\vec\alpha_2\cdot\partial\vec\phi)^{2}
-{1\over \sqrt 2}\,\vec\alpha_1\cdot\partial^2\vec\phi\cr
&-{1\over 2}\, (\vec\alpha_1\cdot\Psi)
(\vec\alpha_1\cdot\partial\vec \Psi)
-{1\over 2}\, (\vec\alpha_2\cdot\Psi)
(\vec\alpha_2\cdot\partial\vec \Psi)\cr\cr
T_F^{\parallel}\,\,=&\,\,-  {1\over 2}
(\vec\alpha_1\cdot\partial\vec\phi)
(\vec\alpha_1\cdot\vec\Psi)
- {1\over 2}\,(\vec\alpha_2\cdot\partial\vec\phi)
(\vec\alpha_2\cdot\vec\Psi)+
{1\over \sqrt 2}\,\vec\alpha_1\cdot\partial\vec\Psi.\cr}
\eqn\susytdos
$$
These parallel components can be expressed locally in terms of the
ghost fields. In fact, using eqs. \susyveinte\ and \susyvuno\ it is
easy to check that
$$
\eqalign{
T_{\parallel}\,\,=&\,\,
-b\partial c- {1\over 2}\, (\beta\partial \gamma
-\gamma\partial \beta)\cr
T_F^{\parallel}\,\,=&\,\,{1\over 2}\,b\gamma
-{1\over 2}\,\beta\partial c,\cr}
\eqn\susyttres
$$
where normal ordering is understood. Denoting the components of $\vec
\phi$, $\vec \Psi$ and $\vec A$ in the orthogonal complement of the
ghost plane as:
$$
\vec \varphi=\vec \phi_{\perp}
\,\,\,\,\,\,\,\,\,\,\,\,\,\,\,\,\,
\vec \psi=\vec \Psi_{\perp}
\,\,\,\,\,\,\,\,\,\,\,\,\,\,\,\,\,
\vec a=\vec A_{\perp},
\eqn\susycuatro
$$
we can write the total energy-momentum tensor $T$ and supersymmetry
generator $T_F$ in the form:
$$
\eqalign{
T\,=\,&-{1\over
2}\,\partial\vec\varphi\cdot\partial\vec\varphi+ \vec
a\cdot\partial^2\vec\varphi+ {1\over
2}\vec\psi\cdot\partial\vec\psi -b\partial c- {1\over 2}\,
(\beta\partial \gamma -\gamma\partial \beta)\cr
T_F\,=&\,\,\,{1\over 2} \,\,\partial\vec \varphi\cdot\vec \psi
-\vec a\cdot\partial\vec\psi+
{1\over 2}\,b\gamma
-{1\over 2}\,\beta\partial c.\cr}
\eqn\susytcinco
$$
Notice that $\vec \varphi$, $\vec \psi$ and $\vec a$ live in an
$n$-dimensional space, with $n=N-2$. The fields $\vec \varphi$ and
$\vec \psi$ will be collectively referred to as ``matter" fields.

Let us now find the expression of the other operators appearing in
the STA. Consider, first of all,  the $R_F$ field.
This operator is the generator of the STA with the lowest dimension
and therefore one expects that the number of possible terms appearing
in its expression will be lower that in any other generator. Recall
that $R_F$ is a fermionic operator whose conformal dimension is
${1\over 2}$. With the  fields we have at hand, the only
contributions that could appear in  $R_F$ are a linear combination of
the components of $\vec \psi$  and the products $\beta c$ and $\gamma c$
of the ghost fields. The fact that the product $Q_B(z)\,R_F(w)$ is
non-singular implies that $R_F$ cannot contain a $\gamma c$ term.
Moreover, the coefficient of the
$\beta c$ term can be determined by requiring  the action of $Q(z)$
on $R_F(w)$ to be given by the first equation in \susycatorce. Thus
one is led to:
$$
R_F\,=\,\beta c- \vec r\cdot\vec\psi,
\eqn\susytseis
$$
where $\vec r$ is a numerical vector. From the expression of $R_F$
given in eq. \susytseis\ one can determine the form of $R$ by
applying $T_F$ to $R_F$ and comparing the result with eq. \susydoce:
$$
R\,=\,bc+\beta\gamma+\vec r \cdot\partial \vec \varphi .
\eqn\susytsiete
$$
In order to match the anomalous behaviour of $R_F$ under the
supersymmetry transformation (see the first equation in \susydoce),
the product $\vec a\cdot\vec r$ must be fixed to a particular value
that depends on $d$. Moreover $\vec r\,^2$ can be determined, for
example, from the second equation in \susyquince. Taking eqs.
\susytres\ and \susytreinta\ into account, one gets the following set
of conditions for the scalar products of the vectors $\vec a$ and
$\vec r$:
$$
\vec a\,^2\,=\,{2-n\over 8}
\,\,\,\,\,\,\,\,\,\,\,\,\,\,
\vec a\cdot\vec r\,=\,-{d+1\over 2}
\,\,\,\,\,\,\,\,\,\,\,\,\,\,
\vec r\,^2\,=\,-d.
\eqn\susytocho
$$

One can follow the same procedure to determine the form of $G_B$
from its behaviour under the action of $Q$ and $Q_B$. After some
trial and error, it is rather easy to arrive at the expression:
$$
G_B\,=\,-cT_{F,M}\,+\,{1\over 2}\,
[bc\gamma+ {1\over 2}\,\beta\gamma^2+\beta c \partial c
+(\vec r\cdot \partial \vec\varphi)\,\gamma -
(\vec r\cdot \vec\psi)\,\partial c +d\partial \gamma],
\eqn\susytochobis
$$
where $T_{F,M}$ denotes the matter contribution to $T_F$. It
remains to obtain the form of $G$. As  in the case of
$R$, once $G_B$ is known
the form of $G$ is fixed by the supersymmetry. Indeed, from the
residue of the single pole singularity in the product $T_F(z)G_B(w)$,
one gets:
$$
\eqalign{
G\,=&\, c[T_M+T_{\beta,\gamma}]-cb\partial c
-\gamma T_{F,M}+\gamma [\beta\partial c-{1\over
4}\,b\gamma]\cr +&(\vec r\cdot \vec\partial\varphi)\partial c
+ {1\over 2}\,(\vec r\cdot\psi)\partial \gamma -
{1\over 2}\,(\vec r\cdot \partial\vec \psi)\gamma +
{d\over 2}\, \partial^2 c.\cr}
\eqn\susytnueve
$$
In eq. \susytnueve\ $T_M$ and $T_{\beta,\gamma}$ are
the contributions to the energy-momentum  tensor of the matter fields
and the $(\beta,\gamma)$ system respectively. It is now
straightforward (although in some cases quite tedious) to check
that all the OPE's of the STA are satisfied if $\vec a$ and $\vec r$
verify eq. \susytocho.

Let us now apply the mirror transformation to the realization of the
STA that we have just obtained. Taking into account the form of
$T$, $T_F$, $R$ and $R_F$ (eqs. \susytcinco-\susytsiete) we get:
$$
\eqalign{
T^*\,=\, &-{1\over
2}\,\partial\vec\varphi\cdot\partial\vec\varphi+ \vec
a\,^*\cdot\partial^2\vec\varphi+ {1\over
2}\vec\psi\cdot\partial\vec\psi -2b\partial c+
c\partial b-
{3\over 2}\,\beta\partial \gamma -
{1\over 2}\,\gamma\partial \beta\cr
T_F^*\,=&\,\,\,{1\over 2} \,\,\partial\vec \varphi\cdot\vec
\psi -\vec a\,^*\cdot\partial\vec\psi+
{1\over 2}\,b\gamma -{3\over 2}\,\beta\partial c
-\partial \beta c,\cr}
\eqn\susycuarenta
$$
where, as in \cnueve, $\vec a\,^*=\vec a -\vec r$. By inspecting eq.
\susycuarenta\ we conclude that in this mirror realization the field
$b$, $c$, $\beta$ and $\gamma$ have conformal dimensions $2$, $-1$,
${3\over 2}$ and $-{1\over 2}$ respectively. Therefore they can be
regarded as the ghost fields of the
Ramond-Neveu-Schwarz (RNS) superstring. Notice
that $G^*=b$ and $G_B^*=\beta$, while
$$
R_F^*\,=\,\vec r\,^*\cdot\vec\psi\,-\,\beta c
\,\,\,\,\,\,\,\,\,\,\,\,\,\,\,\,\,\,\,\,\,\,\,\,\,
R^*\,=\,\vec r\,^*\cdot\partial\vec\varphi\,-\,bc-\beta\gamma,
\eqn\susycuno
$$
with $\vec r\,^*=-\vec r$.
The BRST charge $Q^*$ can be written in the suggestive form:
$$
\eqalign{
Q^*\,=&\, c[T_M^*+{1\over 2}\,T_{gh}^*]
-\gamma [T_{F,M}^*+{1\over 2}\,T_{F,gh}^*]-\cr
-&\partial[ c(R_M^*+{1\over 2}\,R_{gh}^*)]
+{1\over 2}\partial [\gamma (R_{F,M}^*+{1\over
2}\,R_{F,gh}^*)] +{d\over 2}\,\partial^2\,c,\cr}
\eqn\susycdos
$$
where the subscripts $M$ and $gh$ denote respectively the matter and
ghost contributions to the operators appearing in eq. \susycdos. The
first two terms in eq. \susycdos\ correspond to the usual BRST
charge of the RNS superstring, whereas the last three contributions
to $Q^*$ are total derivatives that do not contribute to the
integrated BRST charge. The bosonic partner of the BRST charge can
be written in a way similar to eq. \susycdos. Indeed, after some
calculation, one may verify that $Q_B^*$ is given by:
$$
\eqalign{
Q_B^*\,=&\, -c[T_{F,M}^*+{1\over 2}\,T_{F,gh}^*]
-{\gamma \over 2}\,[R_M^*+{1\over 2}\,R_{gh}^*]-\cr
-&{\partial c\over 2}\, [R_{F,M}^*+{1\over 2}\,R_{F,gh}^*]+
\partial [c (R_{F,M}^*+{1\over
2}\,R_{F,gh}^*)] +{d\over 2}\,\partial\,\gamma.\cr}
\eqn\susyctres
$$
We have thus proved that the BRST operator of the RNS superstring
closes the STA. In complete parallel with what occurs in the bosonic
string, one has to improve the standard BRST superstring current in
order to have a closed operator algebra. Notice the close
relationship between the improving terms in eqs. \ciuno\ and
\susycdos. These improving terms depend on the dimension $d$ and on
the $R^*$ and $R_F^*$ currents. To ensure the fulfillment of the
STA, the vectors $\vec a\,^*$ and  $\vec r\,^*$ must satisfy:
$$
(\vec a\,^*)^2\,=\, {10-n\over 8}
\,\,\,\,\,\,\,\,\,\,\,\,\,\,\,\,\,\,\,
\vec a\,^*\cdot \vec r\,^*\,=\,{1-d\over 2}
\,\,\,\,\,\,\,\,\,\,\,\,\,\,\,\,\,\,\,
(\vec r\,^*)^2\,=\,-d.
\eqn\susyccuatro
$$

As  announced in section 1, the mirror transformation can be
regarded as a transformation of the fields. In order to see how this
can be achieved, let us come back to the fields $\vec \phi$ and
$\vec \Psi$ from which we extracted our ghost system. In terms of
these fields, the $R$ and $R_F$ currents can be written as:
$$
R\,=\, \vec R\cdot\partial\vec \phi
\,\,\,\,\,\,\,\,\,\,\,\,\,\,\,\,\,\,\,\,\,\,\,\,
R_F\,=\, \vec R_F\cdot\vec \Psi,
\eqn\susyccinco
$$
where the numerical vectors $\vec R$ and $\vec R_F$ can be obtained
by evaluating the normal-ordered products appearing in eq.
\susycuno. A short calculation shows that:
$$
\vec R\,=\,-\vec R_F\,=\,\vec r+\vec \lambda.
\eqn\susycseis
$$
Notice that the vectors $\vec R$ and $\vec R_F$ are collinear. Let us
denote by $\vec \phi\,^*$ and $\vec \Psi\,^*$ the fields obtained
after a reflection with respect to a hyperplane orthogonal to $\vec
R$. This reflection is well-defined for $d\not= 0$ (see eq.
\cicuatro). The relation between $\vec \phi$ and $\vec \phi\,^*$ is
given in eq. \cicinco, whereas $\vec \Psi\,^*$ is given by a
similar expression:
$$
\vec \Psi\, ^{*}=\,\, \vec \Psi+ {2\over d}\,
(\vec \Psi\cdot \vec R)\,\, \vec R.
\eqn\susycsiete
$$
In terms of these reflected fields, $T$ and $T_F$ can be written as:
$$
\eqalign{
T=&-{1\over 2}\partial\vec\phi\,^*
\cdot\partial\vec\phi\,^*+
\vec A\,^*\cdot\partial^2\vec\phi\,^*+
{1\over 2}\vec\Psi\,^*\cdot\partial\vec\Psi\,^*\cr
T_F=&\,\,{1\over 2} \,\,\partial\vec \phi\,^*
\cdot\vec \Psi\,^*-\vec A\,^*\cdot\partial\vec\Psi\,^*,\cr}
\eqn\susycocho
$$
where, as in eq. \cisiete, $\vec A\,^*\,=\vec A\,-\,\vec R$. On the
other hand, using eqs. \susyvcuatro, \susyvcinco\ and \susycseis, one
can immediately prove that:
$$
\vec A\,^*\cdot\vec\mu\,=\,{3\over 2},
\eqn\susycnueve
$$
which implies that the conformal dimensions of the vertex operators
$e^{\pm \vec \mu\cdot\vec \phi\,^*}$ with respect to $T$ are:
$$
\Delta (e^{\pm \vec \mu\cdot\vec \phi\,^*})\,=\,
\pm\,{3\over 2}.
\eqn\susycincuenta
$$
Let us now substitute in our bosonization formulas of eqs.
\susyveinte\ and \susyvuno\ the fields $\vec \phi$ and $\vec \Psi$ by
their reflected counterparts $\vec \phi\,^*$ and $\vec \Psi\,^*$. The
resulting ghost fields will be denoted by $b^*$, $c^*$, $\beta ^*$
and $\gamma^*$. One has:
$$
\eqalign{
b\,^*\,=&\,-\vec\mu\cdot\vec\Psi\,^*\,\,
e^{\vec \mu\cdot \vec\phi\,^*}\cr
c\,^*\,=&\, -\vec\lambda\cdot\vec\Psi\,^*\,\,
e^{-\vec \mu\cdot\vec\phi\,^*}\cr
\beta\,^*\,=&\,e^{\vec \mu\cdot \vec\phi\,^*}\cr
\gamma\,^*\,=&\,[\,\vec\lambda\cdot\partial\vec\phi\,^*\,-\,
(\vec\mu\cdot\vec\Psi\,^*)\,(\vec\lambda\cdot\vec \Psi\,^*)]\,\,
e^{-\vec \mu\cdot\vec\phi\,^*}.\cr}
\eqn\susyciuno
$$
After a short calculation one can verify that, if
$\vec \lambda$ and $\vec \mu$  satisfy eq. \susyvcuatro, these ghost
fields have the same OPE's than those of $b$, $c$, $\beta $
and $\gamma$ (see eq. \susyvtres). Notice that, using the result
\susycincuenta, the conformal weights of the ghost fields are:
$$
\Delta (b^*)\,=\,2
\,\,\,\,\,\,\,\,\,\,\,
\Delta (c^*)\,=\,-1
\,\,\,\,\,\,\,\,\,\,\,
\Delta (\beta^*)\,=\,{3\over 2}
\,\,\,\,\,\,\,\,\,\,\,
\Delta (\gamma^*)\,=\,-{1\over 2}.
\eqn\susycidos
$$
Moreover, denoting by $\vec\varphi\,^*$ and $\vec\psi\,^*$  the
components of $\vec \phi^*$ and $\vec \Psi^*$ in the orthogonal
complement of the plane spanned by $\vec \mu$ and $\vec \lambda$, it
is straightforward to prove that $T$ and $T_F$ can be written as:
$$
\eqalign{
T\,=\,&-{1\over
2}\,\partial\vec\varphi\,^*\cdot\partial\vec\varphi\,^*+ \vec
a\,^*\cdot\partial^2\vec\varphi\,^*+ {1\over
2}\vec\psi\,^*\cdot\partial\vec\psi\,^* \cr
&-2b^*\partial c^*+c^*\partial b^*- {3\over 2}\,
\beta^*\partial \gamma^* -{1\over 2}\,\gamma^*\partial
\beta^*\cr\cr
T_F\,=&\,\,\,{1\over 2} \,\,\partial\vec
\varphi\,^*\cdot\vec \psi\,^* -
\vec a\,^*\cdot\partial\vec\psi\,^*+
{1\over 2}\,b^*\gamma^*
-{3\over 2}\,\beta^*\partial c^*-\partial\beta^*c^*.\cr}
\eqn\susycitres
$$
Similarly, one can reexpress $R_F$ and $R$ in terms of the new
fields. One has:
$$
R_F\,=-\beta^*c^*\,-\,\vec r\,^*\cdot\vec\psi\,^*
\,\,\,\,\,\,\,\,\,\,\,\,\,\,\,\,\,\,\,\,\,
R\,=\,-b^*c^*\,-\,\beta^*\gamma^*\,+\vec
r\,^*\cdot\partial\vec\varphi\,^*. \eqn\susycicuatro
$$

It is important to point out the close relationship between eq.
\susycicuatro\ and the second equation in \socho. In both cases the
ghost contributions to the topological currents change their signs
when they are written in terms of the mirror variables whereas in
the matter part $\vec r$ must be substituted by
$\vec r\,^*=-\vec r$. The BRST charge $Q^*$ and its supersymmetric
partner $Q_B^*$ that close the STA with the currents given in eq.
\susycicuatro\ are obtained   by
substituting in eqs. \susycdos\ and \susyctres\
$T^*$, $T_F^*$, $R^*$ and
$R_F^*$ by  $T$, $T_F$, $R$ and $R_F$ respectively. The
expressions for the latter in terms of the new variables are given in
eqs. \susycitres\ and \susycicuatro. One must also replace in eqs.
\susycdos\ and \susyctres\ the ghost fields $b$, $c$, $\beta$,
and $\gamma$ by their mirror counterparts $b^*$, $c^*$, $\beta^*$,
and $\gamma^*$. On the other hand, the BRST ancestors of $T$ and
$T_F$ in this realization are simply $G=b^*$ and $G_B=\beta^*$.
Notice the complete analogy in the behaviour under the mirror
reflection between  the supersymmetric and non-supersymmetric cases
(see eqs. \socho\ and \snueve).

As we have already mentioned, in the variables labelled by an
asterisk, our system can be regarded as a RNS superstring. For a
critical RNS superstring the number $n$ of components of
$\vec\varphi\,^*$ is $10$ and $\vec a\,^*=0$. A glance at eq.
\susyccuatro\ shows that there is only one  value of $d$ consistent
with the closure of the STA:
$$
d\,\,({\rm critical\,\,\, RNS\,\,\,superstring})=1.
\eqn\susycicinco
$$
For this value of $d$, one gets from eq. \susyccuatro\ that $\vec
r\,^*$ cannot be zero since $(\vec r\,^*)\,^2=-1$.

If we were dealing with a non-critical string, one of the directions
of the $n$-dimensional field space must be identified with the
Liouville mode. Accordingly we decompose in this case $\vec a\,^*$,
 $\vec r\,^*$ and $\vec\varphi\,^*$ as in eq. \stuno, while the
Majorana field $\vec \psi\,^*$ is split as:
$$
\vec \psi\,^*\,=\,\vec \psi_m^*\,+\vec \psi_L^*.
\eqn\susyciseis
$$

The central charges $c_m^*$ and $c_L^*$ of the matter and Liouville
degrees of freedom can be computed from their background charges
$a_m^*$ and $a_L^*$ as in section 3. They are now constrained to
satisfy $c_m^*+c_L^*\,=\,15$. As is customary in the study of
superconformal field theories, let us introduce the quantities
$\hat c_m\,=\,{2\over 3}\,c_m$ and
$\hat c_m^*\,=\,{2\over 3}\,c_m^*$. The dimension $d$ of the STA
can be written as a function of $\hat c_m$, $\hat c_m^*$ and the
component of $\vec r$ parallel to the matter subspace. A short
calculation, similar to the one performed in section 3 for the
bosonic string, gives the result:
$$
d\,\,=\,\,{{\hat c_m+\hat c\,^{*}_m\over 2}-\,5\,-\,4\,(\vec
r\,^{*}_{m})\,^2 \, \pm\, \sqrt{(1-\hat
c_m)(9-\hat c^{*}_m)}\over 4}.
\eqn\susycisiete
$$
Remarkably enough, the quantity under the square root in eq.
\susycisiete\ involves the central charges of the matter coupled to
supergravity ($\hat c_m^*$) and of its mirror image ($\hat c_m$). If
we require that $(\vec r\,^{*}_{m})\,^2$ and $d$ be real, the range
of the  allowed central charges is
$\{\hat c_m\leq 1 \,\,\,\,\,\,\,\, \hat c^{*}_m\leq 9\}$ and
$\{\hat c_m\geq 1 \,\,\,\,\,\,\,\, \hat c^{*}_m\geq 9\}$. If $\vec
r\,^{*}_{m}=0$, eq. \susycisiete\ simplifies since in this case
$c_m=c^{*}_m$ and, therefore, the range of values of $c^{*}_m$ that
correspond to real values of $d$ is
$\{ \hat c^{*}_m\leq 1\} \cup \{\hat c^{*}_m\geq 9\}$.

When the number $n$ of components of the fields $\vec\varphi\,^*$ and
$\vec \psi\,^*$ is equal to two, one can solve explicitly the
constraints imposed by the topological symmetry( eq. \susyccuatro).
Notice that the  $n=2$ theories include as particular
cases the minimal models coupled to two-dimensional supergravity
and the $\hat c\,=\,1$ RNS superstring. From eq. \susyccuatro\ we
get that the background charges along the matter and Liouville
directions satisfy $(a_m^*)^2\,+\,(a_L^*)^2\,=\,1$. It is convenient
to parametrize $a_m^*$ and $a_L^*$ as follows:
$$
a\,^{*}_{m}={i\over 2}\,(\,\hat\lambda\,-\,{1\over
\hat\lambda}\,)
\,\,\,\,\,\,\,\,\,\,\,\,\,\,\,\,\,\,\,\,\,\,
a\,^{*}_{L}={1\over 2}\,(\,{1\over \hat\lambda}+\hat\lambda).
\eqn\susyciocho
$$
The quantity $\hat \lambda$ can be determined from the central charge
of the matter sector. After a straightforward calculation one gets:
$$
\hat\lambda=\,\,\sqrt{{9-\hat c\,^{*}_{m}\over 8}}\,\pm\,
\sqrt{{1-\hat c\,^{*}_{m}\over 8}}.
\eqn\susycinueve
$$
In terms of $\hat \lambda$, one can easily get the form of $\vec
r\,^*$ as a function of $d$:
$$
r\,_m^*\,=\,-{i\over 2}\,({1\over
\hat\lambda}\,+\,d\,\hat\lambda\,)
\,\,\,\,\,\,\,\,\,\,\,\,\,\,\,\,\,\,
r\,_L^*\,=\,{1\over 2}\,({1\over
\hat\lambda}\,-\,d\,\hat\lambda\,).
\eqn\susysesenta
$$
A simple calculation shows that the vectors $\vec a\,^*$ and
$\vec r\,^*$ given in eqs. \susyciocho\ and \susysesenta\ solve the
constraints \susyccuatro. Notice that eq. \susysesenta\ determines
the direction of the topological $U(1)$ symmetry in the space of
fields. It is also possible to obtain the central charge $\hat c_m$ of
the matter sector of the mirror theory as a function of $d$ and
$\hat c_m^*$. A short calculation shows that:
$$
\hat c_m\,=\,1-{(d+1)^2\over 2}\,
[\,5\,-\,\hat c\,^{*}_{m}\pm
\sqrt{(1-\hat
c^{*}_m)(9-\hat c^{*}_m)}\,].
\eqn\susysuno
$$
For $d=- 1$ the central charge $\hat c_m$ is equal to one
for all values of $\hat c^{*}_m$, in complete
analogy to what happened in the bosonic case. Indeed, for this value
of $d$,  $a_m^*\,=\,r_m^*$ and $a_L^*\,=\,r_L^*$, which means that the
background charges $a_m$ and $a_L$ are zero. For $d\not=-1$
$\,\,\hat c_m$ depends on $\hat c_m^*$ and it is possible to invert
eq. \susysuno\ and get $\hat c_m^*$ as a function of $d$ and
$\hat c_m$.

\chapter{Summary and Concluding Remarks}

In this paper we have analyzed the possible realizations of the
topological conformal symmetry of CFT's with vanishing Virasoro
anomaly. We have investigated the consequences of the existence of
the mirror symmetry,  which appears quite naturally in our approach
as a transformation that relates two realizations of the topological
algebra. We have seen that, once a convenient basis of fields is
chosen, the mirror automorphism can be recast as a field
transformation.

When the system under study is a model of matter coupled to
two-dimensional (super)gravity, the mirror symmetry seems to play a
relevant role. The BRST current for these theories is just an
improved version of the standard BRST current that fixes the
two-dimensional (super)diffeomorphism invariance.
By means of the mirror transformation one relates
the original model of matter plus two-dimensional (super)gravity to
a system with spin-one ghosts and, what is more important, the
implementation of the topological symmetry depends on the matter
content of the two theories connected by the symmetry. This fact
appears neatly in the expressions that give the dimension $d$  in
terms of the central charges of the matter sector (eqs. \stnueve\
and \susycisiete).

Our results imply that for these (super)gravity theories there
exists an additional BRST symmetry that, apart from a sign, shares the
$U(1)$ currents with the original (super)diffeomorphism invariance.
In general the generator of this new BRST transformation cannot be
expressed locally in terms of the original fields of the model. Only
after the (super)diffeomorphism
ghosts are conveniently bosonized, can a
local expression for the new BRST current  be given. Therefore
one can consider the mirror BRST current as a hidden symmetry of the
matter-plus-gravity models in two dimensions.

It is interesting to connect our analysis with the standard BRST
approach to string theory. In string theory the states are obtained
from the cohomology classes of the BRST charge that satisfy an
additional equivariance condition that involves the zero mode of the
$b^*$ ghost
\REF\Opformalism{L. Alvarez-Gaume, C. Gomez, G. Moore and C.
Vafa\journal\np&B303(88)455; C. Vafa \journal\pl&B190(87)47.}
\REF\ND{P. Nelson\journal\prl&62(89)993; J. Distler and P.
Nelson\journal\cmp&138(91)255.}[\Opformalism, \ND].
This equivariance condition can be understood as a
requirement of covariance of the physical states under general
coordinate transformations of the world sheet
\REF\BCI{C.M. Becchi, R. Collina and C.
Imbimbo\journal\pl&B322(94)79.}[\BCI].
Notice that the $b^*$
field is nothing but the $G$ operator in our realization of the
topological algebra.

The role of the $G$ operator is also crucial in the topological
string theory approach
\REF\Twodgrav{E. Witten\journal\np&B340(90)281;E. Verlinde and H.
Verlinde\journal\np&B348(91)547.}
\REF\KLi{K. Li \journal\np&B354(91)711\journal\np&b354(91)725.}
[\dij, \Twodgrav, \KLi]. In this formalism one couples topological
matter to topological gravity
\REF\LPW{J.M.F. Labastida, M. Pernici and E.
Witten\journal\np&B310(88)611.}[\LPW].
As  has been studied in ref.
\REF\Llatas{J.M.F. Labastida and P.M.
Llatas\journal\np&379(92)220.}[\Llatas],
the coupling of matter to topological gravity can be generated by
means of a gauge principle. The generator of this new symmetry is
precisely $G$ and the corresponding transformations are odd
analogues of the world sheet reparametrizations (recall that $G$ is
the BRST partner of the energy-momentum tensor). In order to have a
theory invariant under the local $G$-transformations, one must
introduce an odd gauge field $\Psi_{\mu\nu}$, which is related to
the world sheet metric $g_{\mu\nu}$ by means of the $Q$ symmetry.
After this odd symmetry is fixed, the standard description of
topological gravity is obtained. It can be shown
\REF\Kanno{T. E. Eguchi, H. Kanno, Y. Yamada and S.-K.
Yang\journal\pl&B305(93)235.}
[\Kanno] that (at
least for minimal models) this topological approach gives rise to
the same physical states as those obtained from the equivariant
cohomology of the standard BRST charge.

To summarize the situation one can say that in both approaches to
 string theory one has to require additional conditions to the BRST
Virasoro cohomology. In the standard approach to  string theory
one considers the equivariant cohomology whereas in the topological
string approach one enlarges the BRST symmetry in such a way that
the equivariance condition is already incorporated in the
topological symmetry. From this point of view  the extra conditions
are generated when the  topological symmetry is
implemented locally since only in this case the field $G$
responsible for these conditions is uniquely determined.
Notice that this is consistent with
what we have obtained since our basic requirement is precisely the
fulfillement of the conditions generated by the topological algebra.
Moreover we have obtained an extra BRST charge in (super)string
theory, different from the standard (super) Virasoro one but closely
related to it, which is obtained from the $G$ operator
by substituting in $G$ the bosonized fields by their reflected
counterparts.

Our results can be generalized in several directions. Let us only
mention that it would be very interesting to study the general form
of the  BRST algebra for the $N\,=\,2$ superstring
\REF\Ademollo{M. Ademollo et
al.\journal\pl&B62(76)105\journal\np&B111(76)77.}[\Ademollo].
In this case, by
a simple counting of the central charges of the ghost sector, we
expect that the lower and upper limits of the barrier will collapse
and, at least for  some choice of the improving terms, there will be no
barrier at all. An interesting aspect to elucidate
is the role played by the mirror symmetry in
the implementation of the topological algebra, which in this case
is a twisted $N\,=\,4$ superconformal algebra as  has been checked
in refs.
\REF\Gomis{J. Gomis and H. Suzuki\journal\pl&B278(92)266;
 A. Giveon and M. Rocek\journal\np&B400(93)145.}
\REF\Boresch{A. Boresch, K. Landsteiner, W. Lerche and A.
Sevrin\journal\np&B436(95)609.}[\Gomis, \Boresch]. Another interesting
question is the relation (if any) of this symmetry with the target
space-world sheet duality that these  $N\,=\,2$ strings have
\REF\OV{H. Ooguri and C.
Vafa\journal\np&B361(91)469\journal\np&B367(91)83
\journal\mpl&A5(90)1389.}[\OV].
Work in these directions is in
progress and we expect to report on it in a near future.

\ack

We would like to thank O. Alvarez, J. M. F. Labastida, M. Mari\~no, J.
Mas and G. Sierra for valuable discussions at different stages of this
work. We are grateful to J. M. Isidro, P.M. Llatas and S. Roy for a
critical reading of the manuscript. This work was supported in part by
DGICYT under grant PB 93-0344 and by CICYT under grant AEN 94-0928.

\endpage

\Appendix A

In this appendix we complete the analysis of the realizations of the
topological symmetry. Consider first of all the non-supersymmetric
case of section 2. As we stated in that section, the values $n$ and
$m$ of the depths of $Q$ and $G$  are restricted by the condition
$n+m\leq 3$ (see eq. \dnueve). Apart from the cases $n=0$, $m=2$ and
 $n=2$, $m=0$ studied in section 2, only when $n=0$, $m=0$ and
$n=1$, $m=1$  is it possible to get a consistent representation of
the TA. For these  depths the number of fields is fixed
to some particular values. For example if $(n,m)\,=\,(0,0)$ one is
forced to have only one scalar field. From eq. \docho\ we obtain $\vec
\alpha^2\,=\,-3$ for this case. The explicit form of the generators
can be easily worked out. One gets
\REF\Waterson{G. Waterson\journal\pl&B171(80)77.} [\Waterson]:
$$
\eqalign{
T\,=&\,-{1\over 2}\,(\partial \phi)^2 +{i\over 2\sqrt{3}}\,\,
\partial ^2\phi\cr
Q\,=&\, {1\over \sqrt 3}\,e^{i\sqrt{3}\phi}
\,\,\,\,\,\,\,\,\,\,\,\,\,\,\,\,\,\,\,
G\,=\, {1\over \sqrt 3}\,e^{-i\sqrt{3}\phi}\cr
R\,=&\,{i\over \sqrt{3}}\,\partial\phi.\cr}
\eqn\apuno
$$
Moreover the value of $d$ for the generators of eq. \apuno\ is fixed
$(d={1\over 3})$. It is important to point out that for the
representation \apuno\ the mirror transformation is simply realized
by the field reflection $\phi\rightarrow -\phi$.

For $(n,m)\,=\,(1,1)$ one can  get a representation of the TA only
when the number of fields is $N=3$, whereas the value of $d$ is
arbitrary. One could proceed as in section 2 and study the most
general operators of depth one that can represent $Q$ and $G$. It is
however easier to introduce a spin one anticommuting ghost system
$(b,c)$ which, in terms of the scalar field $\vec \phi$, is given by:
$$
b\,=\,e^{-\vec\alpha\cdot\vec\phi}
\,\,\,\,\,\,\,\,\,\,\,\,\,\,\,\,\,\,\,\,\,\,\,\,\,\,
c\,=\,e^{\vec\alpha\cdot\vec\phi}.
\eqn\apdos
$$

Notice the difference in the signs of the exponents in eqs. \apdos\
and \ttres. As now $\vec A\cdot\vec\alpha\,=\,-{1\over 2}$ and
$\vec\alpha\,^2\,=\,-1$ ( see eqs. \veinte\ and \docho), the fields
$b$ and $c$ have conformal weights $1$ and $0$ respectively. As in
section 2, one can separate in the energy-momentum tensor the
contributions of the $(b,c)$ system and of the components of $\vec
\phi$ orthogonal to $\vec\alpha$. The latter constitute a
two-component scalar field that we shall denote by $\vec \varphi$. The
expression of $T$ in terms of $b$, $c$ and $\vec \varphi$ is the
same as in eq. \cdos. The background charge $\vec a$ of the field
$\vec\varphi$ must satisfy $\vec a\,^2\,=\,0$. Let us write down the
expressions of the generators for $d\not= 1$ and
$\vec a\not = 0$ ( for $d=1$ and/or $\vec a = 0$ similar
expressions are obtained). The form of $Q$ is:
$$
Q\,=\,\partial c\,+\,{2\over 1-d}\,c\,(\vec a\cdot\partial\vec
\varphi),
\eqn\aptres
$$
while $R$ and $G$ are given by:
$$
\eqalign{
R\,=&\,-bc\,+\,\vec r\cdot\partial\vec\varphi\cr
G\,=&\,{1-d\over 2}\,\partial b\,+b\,\vec
a\cdot\partial\vec\varphi\, -\,b\,(R_M+{1\over 2}R_{gh}).\cr}
\eqn\apcuatro
$$
This representation of the TA corresponds to the standard Coulomb
gas construction of the $N\,=\,2$ superconformal algebra (see, for
example,
\REF\Ito{K. Ito\journal\np&B332(90)566.}ref. [\Ito]).
In eq. \apcuatro\ $R_M$ and $R_{gh}$ are the contributions to $R$
of the $\vec \varphi$ scalar and of the $(b,c)$ system respectively.
The sign of this last term in $R$ imply that the field $c$ has
$R$-charge $+1$. The two component vector $\vec r$ satisfies $\vec
r\,^2\,=\,1-d$ (as in eq. \cseis) together with the condition:
$$
\vec a\cdot\vec r\,=\,{1-d\over 2}.
\eqn\apcinco
$$
It is easy to solve these constraints and find a solution for $\vec
r$. If we parametrize the background charge $\vec a$ as
$\vec a\,=\,\lambda\,(1,\pm i)$ (recall that $\vec a\,^2\,=\,0$),
one gets:
$$
\vec r\,=\,(\,\lambda\,+{1-d\over \lambda}\,,\,
\pm i\,(\,\lambda\,-{1-d\over \lambda}\,)\,).
\eqn\apseis
$$
After performing a mirror transformation to this (1,1) realization,
the fields $b$ and $c$ acquire conformal weights $0$ and $1$
respectively. Thus we generate a realization of the TA similar to
the one in eqs. \aptres\ and \apcuatro\ with the roles of $b$ and
$c$ exchanged.

Let us turn now to the supersymmetric algebra. It is useful in this
case to label the different representations of the STA by the
depths of $Q_B$ and $G_B$. In general we can represent these
operators as:
$$
Q_B\,=\,Q_{B,r}\,\,(\,\partial^i\phi,\,\partial^j\Psi)\,
e^{\vec\mu\cdot\vec\phi}
\,\,\,\,\,\,\,\,\,\,\,\,\,\,\,\,\,\,\,
G_B\,=\,G_{B,s}\,\,(\,\partial^i\phi,\,\partial^j\Psi)\,
e^{-\vec\mu\cdot\vec\phi},
\eqn\apsiete
$$
where $Q_{B,r}$ and $G_{B,s}$ are polynomials in the derivatives of
$\vec \phi$ and $\vec\Psi$ with conformal weights $r$ and $s$
respectively. An analysis similar to the one performed in the
bosonic case shows that now the depths $r$ and $s$ are restricted by
the condition $r+s\,\leq \,2$ (notice that $r$ and $s$ can be half
integers). The realizations of the STA studied in section 4 (\ie\ the
RNS superstring and its mirror) correspond to $(r,s)\,=\,(0,2)$ and
$(r,s)\,=\,(2,0)$. The study of the possible
values of $(r,s)$ allowed by the depth rule yields
a result which is very similar to the bosonic case. In fact only for
$(r,s)$ equal to $(0,0)$ and $(1,1)$ does one arrive at a  consistent
realization of the topological algebra. For $(r,s)\,=\,(0,0)$ the
theory must contain a scalar field and a Majorana fermion. The
dimension is fixed to the value $d\,=\,{1\over 2}$ and the
generators of the STA take the form
\REF\SS{A. Schwimmer and N. Seiberg\journal\pl&B184(87)191.} [\SS]:
$$
\eqalign{
T\,=&\,-{1\over 2}\,(\partial \phi)^2\,+\,
{i\over 2\sqrt{2}}\,\partial^2\phi\,+\,
{1\over 2}\,\Psi\partial\Psi
\,\,\,\,\,\,\,\,\,\,\,\,\,\,\,\,\,\,\,
T_F\,=\,{1\over 2}\,\partial\phi\Psi\,-\,{i\over 2\sqrt 2}\,
\partial\Psi\cr
Q\,=&\,{1\over\sqrt{2}}\,\Psi\,e^{i\sqrt{2}\phi}
\,\,\,\,\,\,\,\,\,\,\,\,\,\,\,\,\,\,\,\,\,
\,\,\,\,\,\,\,\,\,\,\,\,\,\,\,\,\,\,\,\,
Q_B\,=\,{i\over 2}\,e^{i\sqrt{2}\phi}\cr
G\,=&\,-{1\over\sqrt{2}}\,\,\Psi\,e^{-i\sqrt{2}\phi}
\,\,\,\,\,\,\,\,\,\,\,\,\,\,\,\,\,\,\,\,\,
\,\,\,\,\,\,\,\,\,\,
G_B\,=\,{i\over 2}\, e^{-i\sqrt{2}\phi}\cr
R\,=&\,{i\over \sqrt{2}}\,\partial\phi
\,\,\,\,\,\,\,\,\,\,\,\,\,\,\,\,\,\,\,\,\,
\,\,\,\,\,\,\,\,\,\,\,\,\,\,\,\,\,
\,\,\,\,\,\,\,\,\,\,\,\,\,\,\,\,\,\,
R_F\,=\,-{i\over \sqrt{2}}\,\Psi.\cr}
\eqn\apocho
$$

Notice that
for the representation of eq. \apocho\ the mirror transformation is
realized as $\phi \rightarrow -\phi$ and $\Psi \rightarrow -\Psi$, in
complete analogy with what happens for the representation of the TA
given in eq. \apuno.

For $(r,s)=(1,1)$ there exists a representation of the STA with
arbitrary $d$ when the number of bosonic and fermionic fields is
four. In this case it is easier to introduce a supersymmetric ghost
system whose relation with the original fields $\vec \phi$ and $\vec
\Psi$ is as in eqs. \susyveinte\ and \susyvuno. The conformal weights of
$\beta$, $\gamma$, $b$ and $c$ are ${1\over 2}$, ${1\over 2}$, $1$ and
$0$ respectively and therefore $T$ and $T_F$ are given by eq.
\susycinco, where $\vec\psi$ and $\vec\varphi$ are two-component
vectors. For $d\not= 1$  and $\vec a\not = 0$ the expressions of $Q_B$
and $Q$ are:
$$
\eqalign{
Q_B=&\,\gamma\,+\,{2\over 1-d}\,c\,\vec a\cdot\vec\psi\cr\cr
Q\,=&\,\partial c \,+\, {2\over 1-d}\,[c\,\vec a\cdot
\partial\vec\varphi+\gamma\vec
a\cdot\vec\psi\,].\cr}
\eqn\apnueve
$$
The two abelian currents $R_F$ and $R$ take the form:
$$
\eqalign{
R_F=&\,-\beta c\,-\,\vec r \cdot\vec\psi\cr\cr
R\,=&\,-bc\,-\,\beta\gamma\,+\,\vec
r\cdot\partial\vec\varphi,\cr}
\eqn\apdiez
$$
where $\vec r$ is a two-component vector such that $\vec r\,^2\,=-d$
and whose scalar product with the background charge $\vec a$ is
given by eq. \apcinco. Denoting by $R_M$, $R_{M,gh}$, $R_{F,M}$ and
$R_{F,gh}$ to the $\vec \varphi$ and ghost contributions to $R$ and
$R_F$, we can write $G_B$ and $G$ as:
$$
\eqalign{
G_B=&\,{1-d\over 2}\,\partial\beta\,
+{b\over 2}\,(R_{F,M}+{1\over 2}\,R_{F,gh})\,-\,
{\beta\over 2}\,(R_{M}+{1\over 2}\,R_{gh})\,+\cr
\,\,+&{\vec a\cdot\vec\psi\over 1-d}\,
[\,db+\beta(R_{F,M}+{1\over 2}\,R_{F,gh})\,]\cr\cr
G\,=&\,{1-d\over 2}\,\partial b \,-
b\,(R_{M}+{1\over 2}\,R_{gh})\,+
{\partial\beta\over 2}\,(R_{F,M}+{1\over 2}\,R_{F,gh})\,-\cr
\,+&{\beta\over 2}\,(\partial R_{F,M}
+{1\over 2}\,\partial R_{F,gh})\,-
{\vec a\cdot\partial\vec\varphi\over 1-d}\,
[\,db+\beta(R_{F,M}+{1\over 2}\,R_{F,gh})\,]\,-\cr
\,-&{\vec a\cdot\vec\psi\over 1-d}\,
[\,d\partial\beta\,+\,\beta\,(R_{M}+{1\over 2}\,R_{gh})\,+
b\,(R_{F,M}+{1\over 2}\,R_{F,gh})\,].\cr}
\eqn\aponce
$$
The realization of the STA displayed in eqs. \apnueve-\aponce\ was
found previously (although with different notations from ours) in ref.
\REF\Fujitsu{A. Fujitsu\journal\pl&B299(93)49.} [\Fujitsu].
Due to the signs of the ghost contributions to $R$,
the mirror transformation applied to this realization gives rise to a
system of ghosts with spins $1$ and ${1\over 2}$ and with the ghost
number reversed. Actually $b$, $c$, $\beta$ and $\gamma$ acquire
conformal weights $1$, $0$, ${1\over 2}$ and ${1\over 2}$ respectively.
The mirror realization is thus of the same type as the original one
with the roles of $b$ and $\beta$ exchanged with those of $c$ and
$\gamma$ respectively. As in the bosonic case one can solve the
constraints for $\vec r$. If we put again $\vec a\,=\,\lambda\,(1,\pm
i)$, we get:

$$
\vec r\,=\,(\,{\lambda d\over d-1}+{1-d\over 4\lambda}\,,\,
\pm i\,(\,{\lambda d\over d-1}-{1-d\over 4\lambda}\,)\,).
\eqn\apdoce
$$

\endpage

\refout
\end